\def\pfm{$P(F > F_m)$} 

\def\rs{R$_{\rm S}$}
\def\Fm{$F_m$}
\def\la{\hbox{\rlap{\raise.3ex\hbox{$<$}}\lower.8ex\hbox{$\sim$}\ }}
\def\ga{\hbox{\rlap{\raise.3ex\hbox{$>$}}\lower.8ex\hbox{$\sim$}\ }}
\def\alphapl{$\alpha_{\rm PL}$} 
\def\alphapl{$\Gamma$} 
\def\redchisq{$\chi^2_{\rm d.o.f}$}
\def\kalpha{K$\alpha$} 
\def\porb{P$_{\rm orb}$}
\def\Magv{M$_{\rm V}$}
\def\aa{A\&A}

\def\rin{R$_{\rm in}$} 
\def\mns{M$_{\rm NS}$}
\def\onee{1E1724--3045} 
\def\gs{GS1826--238}
\def\slx{SLX1735--269} 
\def\ks{KS1731-260} 

\def\ledd{$\rm L_{\rm Edd}$}
\def\Ledd{$\rm L_{\rm \rm Edd}$}
\def\degree{$^\circ$}

\def\ecutoff{E$_{\rm Cutoff}$}

\def\comptt{{\it Comptt}} 
\def\compps{{\it CompPS}} 
\def\comps{{\it CompPS}} 
\def\dbb{{MCD}} 
 
\def\ktin{kT$_{\rm in}$} 
\def\tin{kT$_{\rm in}$}

\def\teff{kT$_{\rm eff}$} 
\def\bb{{BB}}
\def\compst{{\it Compst}}

\def\lx{L$_{\rm 1-20~keV}$} 
\def\lhx{L$_{\rm 20-200~keV}$} 
\def\lc{L$_{\rm crit}$} 
\def\rxte{{\it Rossi X-ray Timing Explorer~}} 
\def\sax{{\it Beppo-SAX}}
 
\def\dof{d.o.f.}  
\def\ergs{ergs s$^{-1}$}
\def\ergscm{ergs s$^{-1}$cm$^{-2}$}

\def\msol{$M_\odot$} 
\def\deg{$^{\circ}$}
\def\ctss{cts s$^{-1}$} 
\def\nuqpo{$\nu_{\rm QPO}$} 
\def\nuhfqpo{$\nu_{\rm HFQPO}$} 
\def\nub{$\nu_{\rm Break}$} 
 
\def\nh{N$_{\rm H}$} 
\def\nhv{H atoms cm$^{-2}$} 
 
\def\fsoft{F$_{\rm 1-20~keV}$} 
\def\fhard{F$_{\rm 20-200~keV}$} 
\def\dmn{$\times 10^{-9}$ergs s$^{-1}$cm$^{-2}$} 

\def\kte{kT$_{\rm E}$} 
\def\ktw{kT$_W$}
\def\rincost{R$_{\rm in} \sqrt{\cos \theta}$}
 
\def\ktbb{$kT_{\rm BB}$} 
\def\rbb{$R_{\rm BB}$} 
\def\fbb{$f_{\rm BB}$} 
\def\reff{$R_{\rm eff}$} 
\def\rin{$R_{\rm in}$} 
\def\rcol{$R_{\rm in}$} 
\def\rc{$R_C$}

\def\sigfe{$\sigma_{\rm Fe}$}

\def\pexrav{{\sl Pexrav}} 
\def\pexriv{{\sl Pexriv}}
\def\chisq{$\chi^2$} 
\def\relref{f$_{\rm refl}$} 
\def\refrel{f$_{\rm refl}$}

\documentclass[preprint2]{aastex}
\usepackage{psfig}
\shorttitle{Hard X-ray emission from LMXBs}
\shortauthors{Barret et al.}
\begin{document}
 \small
\title{Hard X-ray Emission From Low Mass X-ray Binaries} \author{D.
  Barret, J. F. Olive \& L. Boirin} \affil{\small Centre d'Etude Spatiale des
  Rayonnements, CNRS/UPS, 9 Avenue du Colonel Roche, 31028 Toulouse
  Cedex 04, F} \email{\small Didier.Barret@cesr.fr}
\author{C. Done} \affil{\small Department of Physics, University of Durham,
  South Road, Durham DH1 3LE, England, UK}
\email{\small chris.done@durham.ac.uk}
\author{G. K. Skinner} \affil{\small School of Physics and Astronomy,
  University of Birmingham, Edgbaston, Birmingham B15 2TT, UK}
\email{\small gks@star.sr.bham.ac.uk}
\author{J. E. Grindlay} \affil{\small Harvard Smithsonian Center for
  Astrophysics, 60 Garden Street, Cambridge, MA 02138, USA}
\email{\small josh@head-cfa.harvard.edu}
\begin{abstract}
 \small
  We report on {\it Rossi X-ray Timing Explorer} observations of four
  type I X-ray bursters; namely \onee, \gs, \slx~and \ks. The first
  three were in a low state, with 1-200 keV X-ray luminosities in the
  range $\sim 0.05-0.1$ \Ledd~(\Ledd: Eddington luminosity for a
  neutron star $ = 2.5 \times 10^{38}$ \ergs), whereas \ks~was in a
  high state with a luminosity $\sim 0.35$ \Ledd.  The low state
  sources have very similar power spectra, displaying high frequency
  noise up to $\sim 200$ Hz.  For \ks, its power spectrum is dominated
  by noise at frequencies $\la 20$ Hz; in addition a quasi-periodic
  oscillation at 1200 Hz is detected in a segment of the observation.
  The 1-200 keV spectra of the low state sources are all consistent
  with resulting from thermal Comptonization with an electron
  temperature (\kte) around 25-30 keV.  For \ks, the spectrum is also
  dominated by thermal Comptonization, but with a much lower
  \kte~$\sim 3$keV and no significant hard X-ray emission. With the
  exception of \gs, there is an underlying soft component, carrying at
  most $\sim 25$\% of the total 1-200 keV luminosity. For all sources,
  we have detected an iron \kalpha~ line at 6.4 keV (although it is
  weak and marginal in \onee).  A reflection component is present in
  the spectra of \gs\ and \slx, and for both we find that the
  reflecting medium subtends only a small solid angle ($\Omega/2\pi
  \sim 0.15,~0.28$).  The origin of the line and the reflection
  component is most likely to be irradiation of the accretion disk by
  the X-ray source.
  
  We suggest a model in which the region of main energy release, where 
  hard X-rays are produced would be an optically thin boundary layer 
  merged with an Advection Dominated Accretion Flow (ADAF), and would 
  be responsible for the rapid variability observed.  The soft 
  component observed probably represents the unscattered emission from 
  an optically thick accretion disk of variable inner radius.  When 
  the accretion rate increases, the inner disk radius shrinks and the 
  strength of the reflected component and associated iron line 
  increase.  At the same time, the Comptonization region cools off in 
  response to an increased cooling flux from the accretion disk and 
  from the reprocessed/reflected component, thus leading progressively 
  to a quenching of the hard X-ray emission.  If low state NSs accrete 
  via ADAFs, the observation of X-ray bursts, indicating that all the 
  accreting matter actually accumulates onto the NS surface, argues 
  against the existence of strong winds from such accretion flows.
  
  Finally, we discuss two criteria recently proposed to distinguish
  between non-quiescent Black Holes (BHs) and Neutron Stars (NSs), and
  that are not contradicted by existing observations. The first one
  states that when thermal Comptonization is responsible for the hard
  X-ray emission, only BHs have \kte~larger than $\sim 50$ keV.
  However, this criterion is weakened by the fact that there are NSs
  displaying non-attenuated power laws extending up to at least 200
  keV, possibly implying non-thermal Comptonization or thermal
  Comptonization with \kte~larger than 50 keV. The second criterion
  stipulates that only BHs are capable of emitting hard X-ray tails
  with 20-200 keV luminosities $\ga 1.5 \times 10^{37}$\ergs.
\end{abstract}

\normalsize

\section{Introduction}
It is now well established that hard X-ray emission (E $\ga 30$ keV)
from X-ray binaries is not exclusively associated with Black Hole
systems (BHs).  The major breakthrough came with the SIGMA and BATSE
observations, which provided the first unambiguous detections of type
I X-ray bursters (hence Neutron Star systems, NSs) at $\sim 100$ keV
(Barret \& Vedrenne 1994, Barret \& Tavani 1997 and references
therein).  However, due to the moderate sensitivity and low spectral
resolution of these instruments and the lack of simultaneous X-ray
observations, it was impossible to investigate the conditions under
which NSs emit hard X-rays. Similarly, very little could be inferred
about the accretion geometry and the relative contribution of the
potential emitting regions to the total emission (NS surface, boundary
layer, accretion disk, corona). In addition, the hard X-ray data alone
were not good enough to discriminate between the thermal and
non-thermal models that have been put forward to account for this
emission.  Finally, if the emission of hard X-rays is indeed common to
BHs and NSs, there may remain some differences either in the shape of
the hard tails (position of the energy cutoff; e.g.  Tavani \& Barret
1997, Churazov et al. 1995), or in the luminosities that can be
radiated simultaneously in soft and hard X-rays\footnote{In this
  paper, we define soft X-rays, as photons with energy between 1 and
  20 keV (also X-rays), and hard X-rays, as photons with energy
  between 20 and 200 keV.} (Barret et al. 1996, Van Paradijs \& Van
der Klis 1994, Barret \& Vedrenne 1994, Zhang et al., 1996).  All
these potential differences have yet to be quantified precisely.  They
can now be addressed through observations performed by the \sax~and
the \rxte~(RXTE) satellites; the latter combining the Proportional
Counter Array (PCA) and the High Energy X-ray Timing Experiment
(HEXTE) (Bradt et al.  1993).  These two instruments offer for the
first time the possibility of observing NSs with good sensitivity and
excellent timing capabilities simultaneously from 2 to $\sim 150$ keV.

In this paper, we report on the RXTE observations of four type I X-ray
bursters; namely \onee, \gs, \slx~and \ks. All these sources have in
common that they have already been detected up to $\sim 100$ keV at
least once (Barret \& Tavani 1997). The analysis has been made in a
coherent way for all sources to allow a reliable and consistent
comparison of their respective properties. The paper is organized as
follows. First we review the relevant spectral observations of the
four sources, and present their long term RXTE All Sky Monitor (ASM)
light curves (Sect 2). We then describe our data reduction scheme,
before presenting the results of our RXTE pointed observations (Sect
3).  In section 4, we discuss the main features of our timing and
spectral results and their implications for our understanding of NS
accretion. Finally, we discuss the most recent observational criteria
that have been proposed to distinguish non-quiescent BHs and NSs on
the basis on their broad band spectral properties.

\section{The sources}
First, let us review briefly the previous observations of these four
sources. We will emphasize X-ray spectral observations, in particular
previous \nh~measurements relevant to the PCA spectral analysis,
existing hard X-ray observations, and finally more general information
about these sources.

\subsection{\onee}
\onee~is the most extensively studied of the four sources discussed
here.  It is located in the globular cluster Terzan 2 in the general
direction of the Galactic center.  Its distance was first estimated to
be 7.7 kpc (Ortolani et al. 1997), and recently revised to 6.6 kpc
(Barbuy et al. 1998).  It is a weakly variable X-ray source, and a
persistent, though variable, hard X-ray source (Goldwurm et al.  1993,
1994).  Details about previous X-ray observations can be found in
Barret et al.  (1999a).  Broad band spectral observations have already
been performed by \sax~(Guainazzi et al.  1998).  They showed that the
1 to 200 keV spectrum could be well fitted by the sum of a soft and a
hard Comptonized component.  The former could be equally fitted by a
single temperature blackbody (BB) of \ktbb=0.6 keV, \rbb=12 km, or a
Multi-Color Disk blackbody (MCD, Mitsuda et al.  1984) with \ktin=1.4
keV, \rincost=2.7 km (at 10 kpc corresponding to \rincost= 2.0 km at
6.6 kpc).  The hard Comptonized component was fitted with a
\comptt~model (Titarchuk 1994) with an electron temperature of $\sim
27$ keV, an optical depth $\sim 3.3$ for a spherical scattering cloud,
and a temperature of 0.6 keV (\ktw) for the seed photons.  The ratio
of the bolometric (0.1-100 keV) luminosity of the soft component to
the total luminosity was 11\% and 36\% for the BB and MCD models
respectively.  However, as pointed out by Guainazzi et al.  (1998),
whereas the radius derived from the BB fit is close to the NS radius,
the very small value inferred for \rin\footnote{The MCD model returns
  \tin~and \rincost~(for an assumed distance), which are the color
  temperature of the inner accretion disk and the projected inner disk
  radius (note that in the case of a pseudo-newtonian disk with
  zero-stress inner boundary conditions around a Schwarzschild BH, the
  actual inner disk radius is 2.73$^{-1}$ times less than the measured
  \rin, see Gierlinski et al.  1999). Correction for spectral
  hardening must be made to \tin~and \rin~to account for the fact that
  the inner disk opacity is dominated by electron scattering.  We use
  a spectral hardening factor f=1.7, a value consistent with the
  source luminosity (Shimura \& Takahara 1995), and determine the
  effective temperature as \teff=\tin/f, and the effective inner disk
  radius as \reff $=f^2$ \rcol = 2.9 \rcol.  For \onee, this means
  that \teff=0.82 keV, \reff=11.4 km for an assumed inclination angle
  of 60\deg~(the source is not a dipper, therefore its inclination
  must be less than 75\deg).  If we further assume that the disk
  terminates at the last stable orbit, then as described in Ebisawa et
  al.  (1994), \rs~(the Schwarzschild radius) is related to \reff~as
  3\rs~$\approx 9$ \mns~km $\propto \frac{3}{5} \eta$\reff~(\mns~mass
  of the neutron star in units of \msol, $\eta < 1$ accounts for the
  decrease of \rincost~ by relativistic effects, see Ebisawa et al.
  1994, here we take $\eta=0.6$ as in Shimura \& Takahara (1995)).
  From this, the low inferred value of \reff~implies a very low mass
  of 0.4 \msol~for the NS (for a 1.4 \msol~NS, \reff~should be 37 km,
  equivalent to \rin~$\sim13$ km).  This does not seem to favor the
  MCD model used to fit the soft component in the \sax~data of \onee.
  However, one has to keep in mind that \rin~is derived from the
  normalization of the MCD model, and therefore would be
  underestimated if some fraction of the disk flux is scattered in a
  corona.  The measured value of \rin~derived from the unscattered
  fraction would then depend on the geometry, and the optical depth of
  the scattering corona.}~appears unphysical (for any plausible values
of the source inclination); hence the hypothesis that the soft
component comes from the boundary layer around the NS might be
preferred.

An ASCA observation confirmed the hardness of the source in X-rays,
and allowed an accurate determination of \nh~towards the source ($\sim
1.2 \times 10^{22}$ \nhv).  This value is consistent with that
expected from the optical reddening of the cluster (Barret et al.,
1999a).

\subsection{\gs}
When \gs~was discovered serendipitously by GINGA in 1988, it was first
classified as a transient source, and further classified as a Black
Hole Candidate (BHC) based on its rapid and intense flickering and
hard Power Law (PL) spectrum in X-rays (Tanaka 1989).  The source was
later detected by TTM (In't Zand 1992), ROSAT-PSPC (Barret et al.
1995a), and at hard X-ray energies by OSSE (Strickman et al. 1996).
The BH nature of \gs~was questioned (e.g. Strickman et al. 1996,
Barret et al. 1996), but the firm evidence that \gs~did not contain a
BH came with the discovery of type I X-ray bursts (the unmistakable
signature of a NS) with the \sax~Wild Field Cameras (WFC, Ubertini et
al. 1997).  In the WFC data, the 70 bursts observed over more than 2.5
years of source monitoring exhibit a quasi-periodicity of 5.76 hours
in their occurrence times (Ubertini et al. 1999). In addition, the
so-called $\alpha$ parameter which is the ratio between the average
persistent X-ray flux to the average flux emitted during X-ray bursts
(assuming that both are isotropic) was estimated to be $60 \pm 7$ for
two of the 70 bursts analyzed by Ubertini et al.  (1999). Similarly
In't Zand et al. (1999), using also a sample of two bursts, observed
$\alpha = 54\pm5$ consistent with the previous estimate. This value is
consistent with a picture in which all the accreting material is
accumulated onto the NS and helium is burned during the bursts
($\alpha$ should be $\sim 20$ and $\sim 80$ for pure Hydrogen and
Helium burning respectively).

As is the case for \onee, broad band observations of \gs~already
exist.  First, Strickman et al. (1996), combining non simultaneous
GINGA (X-ray) and OSSE (hard X-ray) data showed that the 2-200 keV
spectrum of the source could be fitted by an exponentially cutoff
power law (CPL: \alphapl$=1.76\pm0.02$, \ecutoff$=58\pm5$ keV,
\alphapl~is the photon index of the power law, i.e. the energy index
would be \alphapl-2) with a weak (and marginally significant)
reflection component. More recently, \sax~observed the source,
following a BATSE trigger which indicated that \gs~was emitting hard
X-rays (Frontera et al. 1998, Del Sordo et al. 1999).  The 0.5-200 keV
spectrum was well fitted by a composite model consisting of a
blackbody of $0.94\pm0.05$ keV, and a PL of photon index
\alphapl$=1.34\pm0.04$, exponentially cutoff at $49\pm3$ keV.  The
column density measured by \sax~was $0.47\times 10^{22}$ \nhv,
consistent with the ROSAT PSPC value ($0.5 \pm 0.04 \times 10^{22}$
\nhv, Barret et al. 1995); both being slightly larger than the value
expected from the optical reddening towards the source (E(B-V)$\sim
0.4$ corresponding to $0.22\times 10^{22}$ \nhv, Predehl \& Smith
1995, Barret et al. 1995). Another \sax~observation performed half a
year before by In't Zand et al.  (1999) yielded best fit spectral
parameters consistent with those reported in Del Sordo et al.  (1999).

\gs~has a V$=19.1\pm0.1$ optical counterpart (Barret et al. 1995,
Homer et al. 1998) in a possible 2.1 hour orbital period system, and
is therefore a Low Mass X-ray Binary (LMXB).  If the 2.1 hour
modulation is the true orbital period of the system, one can try to
estimate the source distance following the approach of Van Paradijs \&
McClintock (1994). In LMXBs, the bulk of the optical light originates
from the reprocessing of the X-rays in the accretion disk.  For
systems for which reliable distance estimate exists, there is a good
correlation between the absolute V magnitude and the X-ray luminosity
($L_X$) and the size of the accretion disk (and hence the orbital
period \porb, $ L_V \propto L_X^{1/2}$ \porb$^{2/3}$).  Most LMXBs
have \Magv~in the range 0-2, whereas some short period systems have
\Magv~in the range 3-5.  With a 2.1 hr orbital period, \gs~is a short
period system, and if one assumes \Magv~in the above range (and an
absolute dereddened V magnitude of 17.9), one gets a formal distance
range between $\sim 4$ and $\sim 10$ kpc. Constraints on the source
distance can also be inferred from the observation of X-ray bursts.
In't Zand et al. (1999) derived an unabsorbed bolometric peak flux of
a burst of $2.7\pm0.5 \times 10^{-8}$ \ergscm.  The fact that this
burst did not show evidence for photospheric expansion implies that
its luminosity is below the Eddington limit (\Ledd) which we assume to
be $2.5 \times 10^{38}$ \ergs~(this value is appropriate for
helium-rich material, a 1.4 \msol~NS, and a moderate gravitational
redshift correction, Van Paradijs \& McClintock 1994).  This in turn
implies an upper limit on the source distance of 9.6 kpc.  In the
remainder of this paper, we assume a distance of 7 kpc for \gs.

\subsection{\slx}
\slx~was reported for the first time in 1985, when it was detected by
the {\it Spacelab 2} X-ray Telescope (Skinner et al. 1987). However,
it was present in the Einstein Slew Survey in observations performed
$\sim 5$ years earlier (Elvis et al. 1992). \slx~was later detected by
TTM (In't Zand 1992), by ART-P, ROSAT-PSPC (Grebenev et al. 1996) and
ASCA (David et al.  1997).  SIGMA observations have shown that it is a
persistent hard X-ray source of the Galactic center region (Goldwurm
et al. 1996).  The weakness of the source ($\sim 15.4$ mCrab in the
35-75 keV range) did not allow tight constraints to be put on the time
averaged (1990-1994) hard X-ray spectrum, which could be fitted either
by a simple PL of photon index \alphapl$\sim 2.9\pm0.3$, or
alternatively by a Comptonization model (\compst~in XSPEC, Sunyaev and
Titarchuk, 1980) with \kte~of $26^{+35}_{-9}$ keV (Goldwurm et al.
1996).  Although early suspected to contain a NS based on the softness
of its hard X-ray spectrum (Goldwurm et al.  1996), its nature
remained uncertain until type I X-ray bursts were discovered with the
\sax~WFC (Bazzano et al. 1997).  The ASCA observations of
\slx~revealed that its 0.6-10 keV spectrum could be well fitted with a
PL of index 2.15 (David et al. 1997), absorbed through an \nh~of $\sim
1.4-1.5 \times 10^{22}$ \nhv, a value which is consistent with a
source location near the Galactic center (i.e. at a distance of $\sim
8.5$ kpc, David et al. 1997). The \nh~value derived with ASCA is also
consistent with the one derived from the ROSAT-PSPC and ART-P
observations ($1.2-1.4 \times 10^{22}$ \nhv, Grebenev et al.  1996).
The rapid variability of the source was recently investigated by
Wijnands \& Van der Klis (1999), using a short 10 kilosecond
\rxte~observation performed between February and May, 1997 (see
below).

No accurate distance estimate exists so far.  From the burst reported
in Bazzano et al. (1997a) ($1.5 \times 10^{-8}$ \ergscm, 2-10 keV,
Bazzano et al. 1997b), an upper limit of $\sim 10$ kpc can be derived
(after bolometric corrections and assuming a blackbody of 2 keV for
the burst). In the remainder of the paper, we will assume a distance
of 8.5 kpc.
 
\subsection{\ks}
\ks~was discovered in October 1988 by TTM (Sunyaev et al. 1990), and
then classified as a transient.  \ks~is located in the Galactic center
region, about 5\deg~away from the Galactic nucleus and only 1\deg~from
the X-ray pulsar GX1+4.  Above the mean persistent level of 80 mCrab
($2.7 \times 10^{-10}$\ergscm, 1-20 keV), TTM observed several X-ray
bursts from \ks, thus indicating that it contains a weakly magnetized
NS. The TTM spectrum was fitted by a Thermal Bremsstrahlung (hereafter
TB\footnote{Note that at such luminosities, TB fits are unphysical
  because they imply emission measures $ 10^{59}-10^{60}$ cm$^{3}$.
  The optically thin requirement leads to a size of $10^{10}-10^{11}$
  cm for the cloud.  These sizes are at least $\sim 100-1000$ times
  larger than the region of main energy release around a NS. This
  means that one should be cautious about the \nh~fitted, because its
  value depends critically on the shape assumed to fit the
  continuum.}) of 5.7 keV, absorbed through an \nh~ of $2.2\times
10^{22}$ \nhv.  Follow-up ROSAT and optical observations indicated
that the source is a likely LMXB (Barret et al. 1998).  The
\nh~derived from the ROSAT-PSPC all sky survey observations of
$1.3\times 10^{22}$ \nhv~for a PL fit; is about a factor of 2 less
that the value derived with TTM, thus possibly indicating intrinsic
\nh~variations within the source.  More recently, we observed \ks~with
ASCA (September 27th, 1997) and confirmed the \nh~found by ROSAT
(Narita et al.  1999).  Additional X-ray observations can be found in
Yamauchi \& Koyama (1990) and Aleksandovich et al.  (1995).  \ks~ was
detected only once in hard X-rays between 35 and 150 keV by SIGMA, and
has remained undetectable since then at these energies (Barret et al.
1992).  Unfortunately, no simultaneous X-ray observations exist for
the SIGMA detection.

RXTE first observed \ks~in July-August 1996. It was then in its
standard high state.  These observations led to the discovery of a
highly coherent 524 Hz periodic X-ray signal at the end of the
contraction phase of an X-ray burst that showed photospheric expansion
(Smith et al. 1997). This signal was tentatively interpreted as an
indication of the NS spin (Smith et al. 1997). However, in the same
data set, two simultaneous High Frequency Quasi-Periodic Oscillations
(HFQPOs) at $898\pm3.3$ Hz and $1158\pm9$ Hz were also found (Wijnands
\& Van der Klis 1997), see below). The frequency separation
($260.3\pm9.6$ Hz, corresponding to 3.8 msec) which is equal to half
the frequency of the burst oscillations was then interpreted as the
true NS spin frequency (Wijnands \& Van der Klis 1997).

In addition, assuming that the burst observed by RXTE reached the
Eddington luminosity, a distance of $8.8\pm0.3$ kpc was derived for
the source (Smith et al.  1997).  At this distance, the implied source
X-ray luminosity at the time of the RXTE observations was $5.7 \times
10^{37}$\ergs.  Finally a \nh~of $6\times 10^{22}$ \nhv~ was derived
for a TB fit of the RXTE PCA spectrum; a high value, significantly
larger than the ASCA, ROSAT and TTM ones.

\subsection{ASM light curves}

\begin{figure}[!h]
  \centerline{\psfig{figure=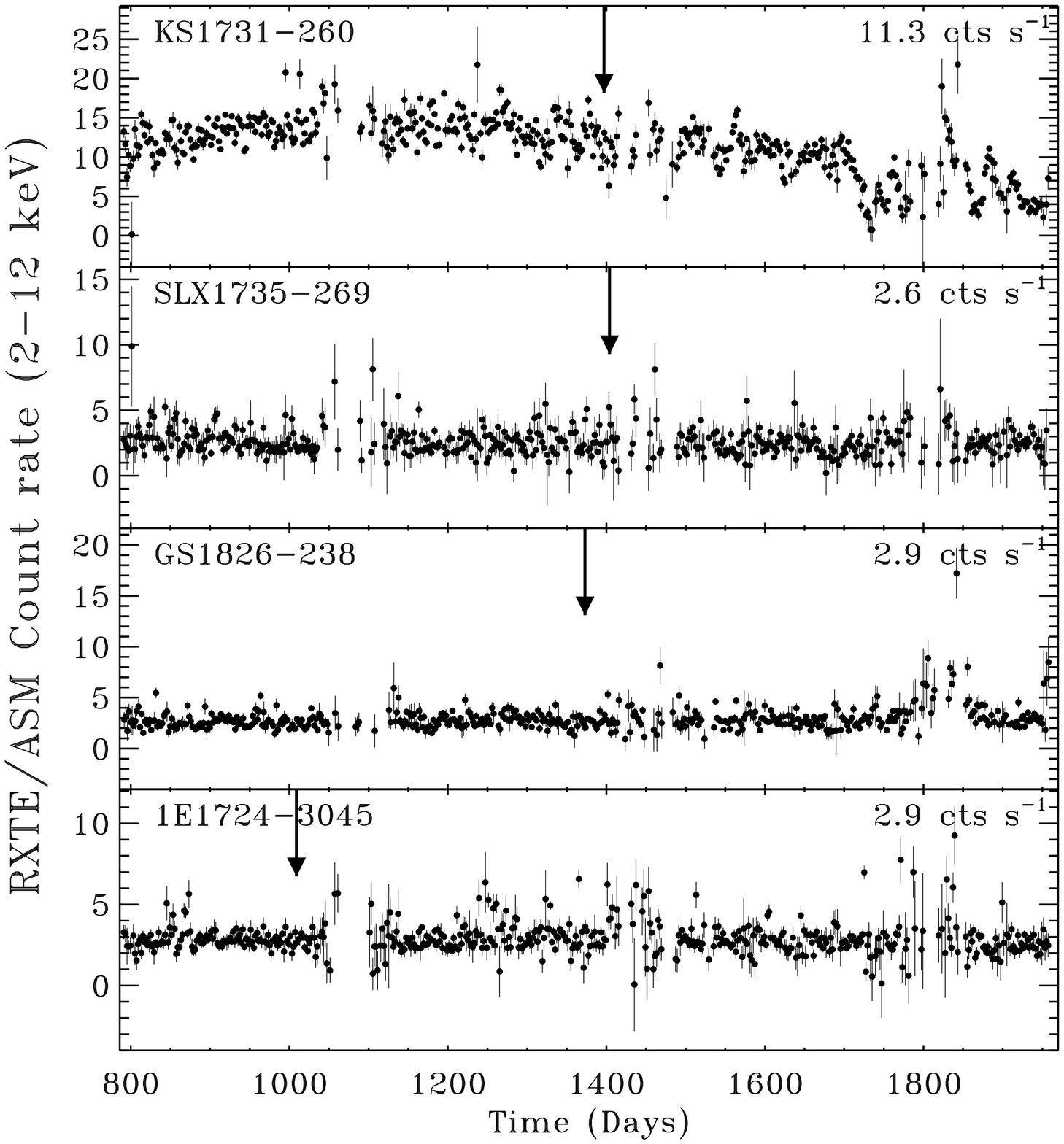,height=8.5cm}}
\caption{The RXTE/ASM long term light curves of \onee, \gs, \slx~and \ks. 
  Time is given as days after January 1st, 1994. Data have been
  selected using {\it fselect} to satisfy the criterion that the
  measured rate for each individual ``dwells'' is positive and that
  the reduced $\chi^2$ of the fit to recover the source intensity is
  less than 1.5. Data from all three Scanning Shadow Cameras are
  combined within {\it lcurve}. The binning time of the light curves
  is 2 days. The gaps in the light curves are due to the sun getting
  too close to the source region (see Levine et al. 1996 for
  information about the ASM). The dates of our pointed observations
  are indicated with arrows.}
\label{lcasm}
\end{figure}

The RXTE ASM long term light curves of all four sources are shown in
Fig.  \ref{lcasm} (for information about the RXTE/ASM see Levine et
al. 1996).  The mean ASM count rates are 2.9 \ctss~($\sim 38$ mCrab),
2.9 \ctss~($\sim 38$ mCrab), 2.6 \ctss~($\sim 35$ mCrab), 11.3
\ctss~($\sim 150$ mCrab) in the 2-12 keV range for \onee, \gs,
\slx~and \ks~respectively. Assuming a PL spectrum of photon index 2,
and the source \nh, at the source distance, these values correspond to
average 1-20 keV luminosities of $\sim 1.0\times 10^{37}$ \ergs$=0.04$
\ledd, $8.8\times 10^{36}$ \ergs$=0.03$ \ledd, $1.3\times 10^{37}$
\ergs$=0.05$ \ledd~for \onee, \gs~and \slx~respectively.  For \ks,
assuming a thermal-like X-ray spectrum, the mean flux is about a
factor of 5 larger: $\sim 4.9 \times 10^{37}$ \ergs~$= 0.2$ \ledd.
These light curves provide the first confirmation of the low
variability of \slx, \gs~and \onee. In particular, for \gs, this
together with independent measurements (e.g. Barret et al. 1995; In't
Zand et al.  1999) indicate that since its discovery in 1988, the
source is a very stable accretor, and certainly not a classical
transient as previously thought. The light curve of \ks, shows also
that the source is persistent but occasionally undergoes low intensity
states, which might be associated with episodes of hard X-ray
emission.

\section{Observations and Results}

The PCA instrument consists of a set of 5 identical Xenon proportional
counters covering the 2-60 keV energy range with a total area of about
6500 cm$^2$ (Bradt et al. 1993). The HEXTE instrument is made of two
clusters of 4 NaI(Tl)/CsI(Na) phoswich scintillation detectors
providing a total effective area of 1600 cm$^2$ in the 15-200 keV
range\footnote{Note that this is about a factor of 4 larger that
  the effective area of the SIGMA imaging telescope for a source
  located in its fully coded field of view (Paul et al. 1991).}
(Rothschild et al. 1998).  The two clusters rock alternately between
the source and background fields to measure the background in real
time. Special care has been taken to avoid the presence of any known
X-ray sources in the background fields.

The rapid variability of the 4 sources has been investigated with the
PCA {\it science event} data. Our spectral analysis is based on the
{\it standard 2} data for the PCA. They provide count spectra each 16
seconds in 128 energy channels covering the 2 to 100 keV range. The
FTOOLS used for PCA background estimation is {\it pcabackest} version
2.1b (released in October 1998). For HEXTE, we also use the standard
mode data.  They provide 64 channel count spectra each 16 seconds.
PCA spectra have been first accumulated for each PCU units for
integration times varying from $\sim 1000$ to $\sim 2000$ seconds,
roughly equivalent to an orbit after filtering for good time
intervals. In particular, PCA data recorded during and up to 30
minutes after the SAA passage have been removed using the FTOOLS 4.2
version of {\it xtefilt}.  PCA response matrices were then computed
for each PCU using the FTOOLS {\it pcarsp 2.36}.  We have then summed
individual PCU spectra using {\it addspec} version 1.0.1, which
combines not only the PHA files, but also the background PHA files,
and computes the associated response matrices.

For HEXTE, spectra were accumulated for each cluster with similar
integration times to those used for the PCA. The latest response
matrices have been used {\it (hexte\_97mar20c\_pwa.rmf)}. Cluster A
and B spectra were then combined together using {\it addspec}, to get
a single HEXTE spectrum averaged over the whole observation. The
source and background light curves and spectra were corrected for
deadtime effects using version 0.0.1 of {\it hxtdead}.

Version 10.00 of XSPEC (Arnaud 1996) has been used for the spectral
fitting.  When combining PCA and HEXTE spectra, the energy ranges of
the fit were 2.5 to 25 keV for the PCA, and above 25 keV and up to at
most 150 keV for HEXTE. These energy ranges were selected to exclude
the steepening of the PCA spectrum above 25 keV (in most spectra, it
starts around 20 keV), as well as the flatening of the HEXTE spectrum
towards low energies.  Starting at 2.5 keV with the PCA implies that
in most cases, the \nh~cannot be very well constrained.  In each case,
we have therefore set \nh~to the most accurate value measured so far
(i.e by ASCA/ROSAT/\sax).  In order to account for the uncertainties
in the relative calibration of the PCA and HEXTE experiments, we have
also left the relative normalization of the two spectra as a free
parameter of the model used.

To investigate the systematics in the PCA data and to validate our
analysis scheme, we have first analyzed a 15 kilo-second Crab
observation (March 22, 1997).  We have found that to get acceptable
fits, it was necessary to add a systematic error to the data to reduce
the effects of the imperfect knowledge of the instrument response near
the K and L edges of xenon.  These errors derived from looking at the
residuals of the power law fit are 0.5\% between 2.5 and 15 keV, 1\%
between 15 and 25 keV and 2.5\% above.  Fitting the Crab spectrum
between 2.5 and 25 keV thus yields a \redchisq~of 1.0 (50 \dof), a
power law index of 2.18, and a normalization at 1 keV of 13.5 Photons
cm$^{-2}$ s$^{-1}$ keV$^{-1}$ (\nh$=3.3\times 10^{21}$\nhv~being
frozen during the fit). The ratio between the data and the power law
folded through the response matrice is shown in Fig. 8. A similar
level of systematics has been used by several groups (e.g. Rothschild
et al.  1999, Wilms et al.  1999, Bloser et al. 1999).  Based on this,
before the fitting, we have combined quadratically to the Poisson
errors of our data, this level of systematics using the FTOOLS {\it
  grrpha} version 5.7.0.  No systematic errors were added to the HEXTE
data.

\subsection{The RXTE observations}

The RXTE observations are summarized in Fig. 2, 3, 4 and 5, for \onee,
\gs, \slx, and \ks. These figures represent the HEXTE
background-subtracted light curve (top panel), the PCA 2-40
background-subtracted light curve (underneath), the PCA color-color
diagram (soft color: 7-15 keV/2-7 keV, hard color 15-40 keV/7-15 keV
(bottom left panel), and a PCA hardness intensity diagram (bottom
right panel).  For clarity and homogeneity, the data from orbits
contaminated by bursts have been filtered out (only data recorded with
the 5 PCA units ON are shown).

The observation of \onee~started on November 4th, 1996 (T0=11h 54mn
39s), and was spread over more than 4 days.  During the observation,
an X-ray burst occurred on Nov.  8th at 05h 30mn 07s (UT).  As said
above for \ks, a few NSs exhibit nearly coherent pulsations during
X-ray bursts.  Sampling the burst profile each second, we have
searched for such a coherent signal in both the {\it science event}
and {\it burst catcher} mode data.  No significant signals were
detected between 200 and 1500 Hz.  Data from the orbit when the burst
occurred and the next one have been removed from the present analysis.
The source did not display any significant variability over the
observation.  In addition, looking at the color-color diagram, one can
see that it was observed in a single spectral state (see also, Olive
et al. 1998). A single PCA spectrum averaged over the whole
observation has thus been considered for the spectral analysis.

\begin{figure}[!th]
\centerline{\psfig{figure=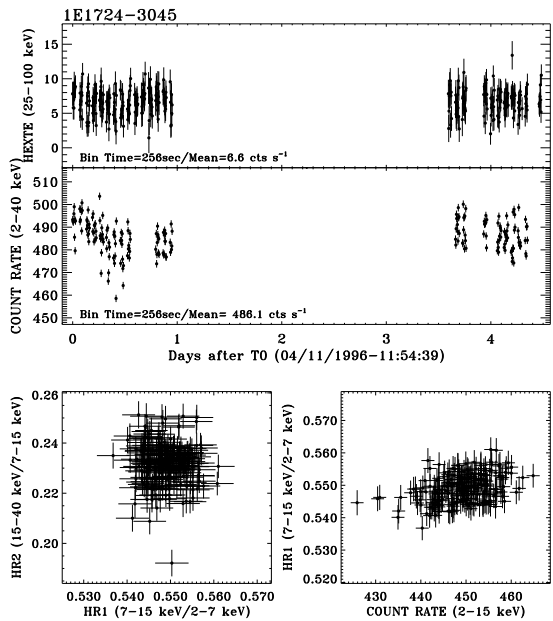,height=8.5cm}}
\caption{{\it Top} HEXTE 25-100 keV  hard X-ray light curve with underneath 
  the 2-40 keV PCA light curve of \onee~over the 4.5 day span of our
  observation. {\it Bottom left} The color-color diagram (the soft
  color HR1 is the ratio 7-15 keV/2-7 keV and the hard color HR2 is
  15-40 keV/7-15 keV) and {\it Bottom right} the hardness-intensity
  diagram for the soft color. Only PCA data recorded with the 5 PCA
  units ON are shown (this explains the first gap in the PCA light
  curve while HEXTE was recording data).}
\label{tz2_summary}
\end{figure}

\begin{figure}[!h]
\centerline{\psfig{figure=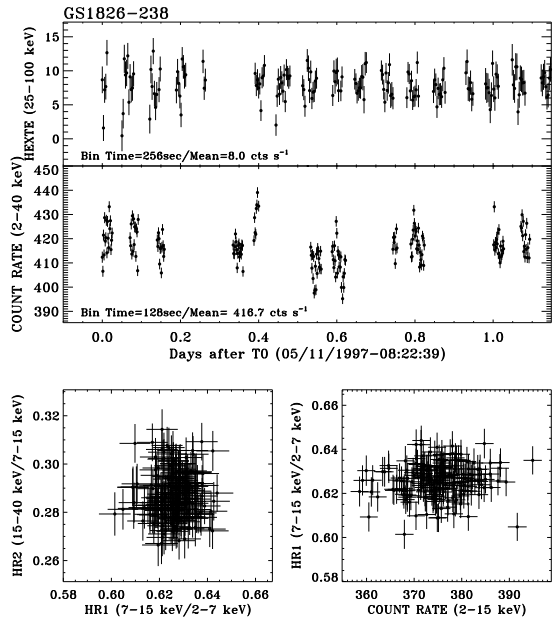,height=8.5cm}}
\caption{{\it Top} HEXTE 25-100 keV  hard X-ray light curve with 
  underneath the 2-40 keV PCA light curve of \gs~over the 1.2 day span
  of our observation. {\it Bottom left} The color-color diagram (the
  soft color HR1 is the ratio 7-15 keV/2-7 keV and the hard color HR2
  is 15-40 keV/7-15 keV) and {\it Bottom right} the hardness-intensity
  diagram for the soft color. The data contaminated by X-ray bursts
  have been removed.  Only PCA data recorded with the 5 PCA units ON
  are shown. }
\label{gs_summary}
\end{figure}

\gs~was observed between November 5th and 6th, 1997 for about 50 ksec
also (T0=08h 22mn 39s, UT).  There are two bursts in the middle of the
observation.  As for \onee, no coherent pulsations were detected in
those bursts.  The time separation between the two bursts is $\sim 5.6
$ hours, consistent with the 5.76 hours (1$\sigma$ spread of 0.26 h)
periodicity reported by Ubertini et al.  (1999).  In addition, from
this periodicity, one expects a burst to have occurred $\sim 1600$
seconds before the beginning of our observation.  The tail of this
burst is clearly seen in the first orbit of data.  Data from this
orbit, as well as from those contaminated by the two bursts, have been
removed from the present analysis.  The spectral analysis of these
bursts and their impact on the persistent emission will be reported
elsewhere.  As shown in Fig.  \ref{gs_summary} the source intensity
varies by as much as $\sim 10$\% in the 2-40 keV range, but the source
did not display significant spectral variability. As for \onee, a
single PCA spectrum was then used in the spectral analysis.  In the
hard X-ray band, it is the brightest of the sources considered here
(the count rate in HEXTE-Cluster A is 8.0 \ctss~in the 25-100 keV
range).

\slx~was observed in October 10th, 1997 (T0=04h 37mn 19s); its
observation which was scheduled for 50 ksec ended 2.5 days later.
The mean PCA count rate is 165.5 \ctss~(2-40 keV), and this is the
faintest of the four sources considered here.  No X-ray bursts were
observed.  The source is clearly variable on minute time scales.
However it is clear that these intensity variations are not
accompanied by strong spectral changes, as the source remains in a
limited region in the color-color diagram.

\begin{figure}[!h]
\centerline{\psfig{figure=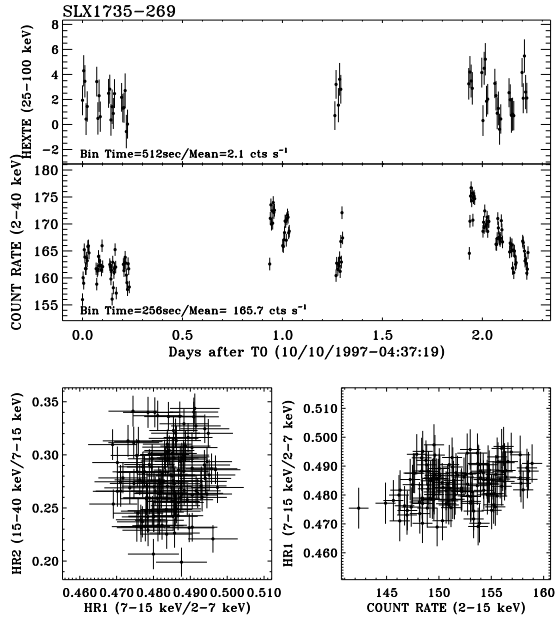,height=8.5cm}}
\caption{{\it Top} HEXTE 25-100 keV  hard X-ray light curve with underneath 
  the 2-40 keV PCA light curve of \slx~over the 2.5 day span of our
  observation. {\it Bottom left} The color-color diagram (the soft
  color HR1 is the ratio 7-15 keV/2-7 keV and the hard color HR2 is
  15-40 keV/7-15 keV) and {\it Bottom right} the hardness-intensity
  diagram for the soft color.  Only PCA data recorded with the 5 PCA
  units ON are shown. }
\label{slx_summary}
\end{figure}

Finally, the observation of \ks~started on October 28th, 1997 (T0=22h
20mn 15s, UT), lasted for about 50 ksec, and ended on October 30th. In
the PCA, \ks~has the largest count rate of the sources considered
here.  The so-called banana shape is clearly visible on the
color-color and hardness intensity diagrams. It moves from the lower
to the upper parts of the banana as the count rate increases (by up to
$\sim 25$\%, 2-40 keV). Given the source variability, we have
considered three PCA spectra in the spectral analysis; one for each
day of the observation.  It is detected by HEXTE only up to $\sim 50$
keV. No X-ray bursts were detected.

\begin{figure}[!h]
\centerline{\psfig{figure=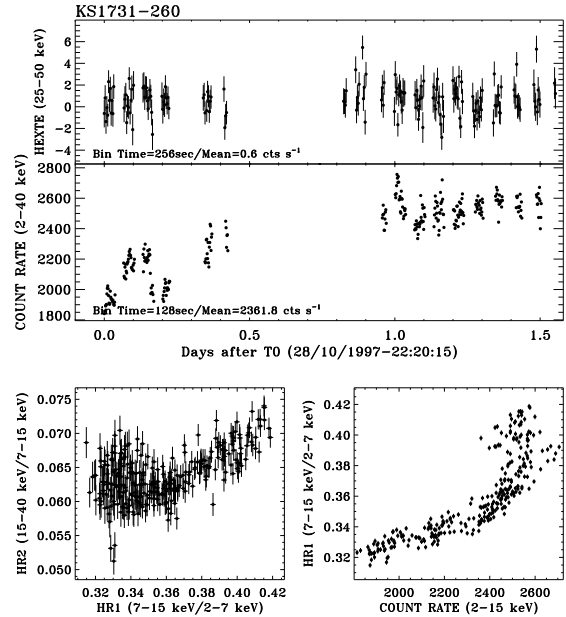,height=8.5cm}}
\caption{{\it Top} HEXTE 25-50 keV  hard X-ray light curve with underneath 
  the 2-40 keV PCA light curve of \ks~over the 1.5 day span of our
  observation. {\it Bottom left} The color-color diagram (the soft
  color HR1 is the ratio 7-15 keV/2-7 keV and the hard color HR2 is
  15-40 keV/7-15 keV) and {\it Bottom right} the hardness-intensity
  diagram for the soft color.  Only PCA data recorded with the 5 PCA
  units ON are shown. }
\label{ks_summary}
\end{figure}

\subsection{Timing properties}
\begin{figure}[!th]
\centerline{\psfig{figure=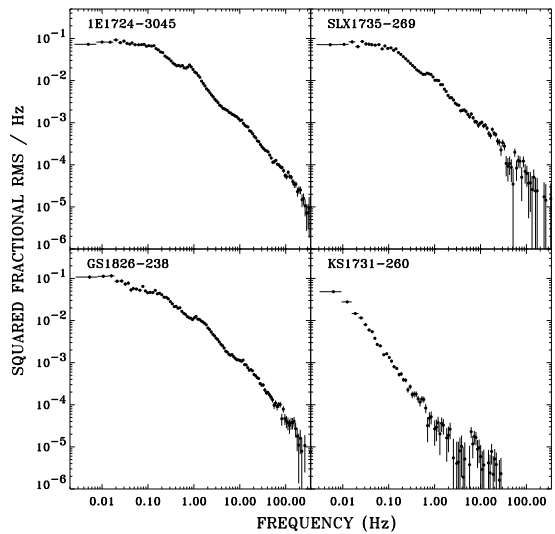,height=7.5cm}}
\caption{The power density spectra of all four sources averaged over
  the whole observations. \onee, \gs~and \slx~display high frequency
  noise extending up to 200-300 Hz with RMS amplitudes of 29.1\%,
  26.1\%, 27.6\% in the 2-40 keV band (0.005-300 Hz). \ks~displays
  Very Low Frequency Noise with a RMS amplitude of $\sim 3.0$\% (2-40
  keV). These PDS are consistent with the first three sources being in
  the so-called island state (low state), and \ks~being in the banana
  state (high state).}
\label{4pds}
\end{figure}

The normalized Power Density Spectra (PDS) averaged over the whole
observation of all four sources are shown in Fig.  \ref{4pds}.  The
first three sources show noise up to $\sim 200$ Hz.  The integrated
power of these PDS is 29.1\%, 26.1\%, 27.6\% in the 2-40 keV band
(0.005-300 Hz).  In addition, they also show a break in the PDS at low
frequencies (typically around 0.1-0.2 Hz), and a QPO-like feature
around 1 Hz.  Although the complete analysis is still under way (for
\gs~and \slx), no strong HFQPOs were detected from any of these three
systems.  For \onee, for which a sensitive search has already been
conducted, an upper limit of 2.5\% on the RMS of a 1000 Hz QPO has
been derived (5-30 keV) (Barret et al.  1999b).

The PDS shown in Fig.  \ref{4pds} for \slx~is very similar both in
shape and normalization to the one reported by Wijnands \& Van der
Klis (1999a) from a short RXTE observation performed in February-May
1997.  Although the source count rate did not change a lot during our
observation (see Fig.  4), we have computed two PDS for segments of
the observation associated with the highest and lowest count rates:
172 \ctss~(156 \ctss) versus 162 \ctss~(150 \ctss), 2-40 keV (2-16
keV), respectively.  Fitting these two PDS with broken power laws
yields a break frequency of $0.15\pm0.1$ Hz, and $0.08\pm0.02$ Hz
respectively.  Thus, within our limited range of intensity variations,
the break frequency decreased with the count rate, a behaviour which
is generally observed from atoll sources (Van der Klis, 1994).  At
lowest count rate, Wijnands \& Van der Klis (1999a) reported the
opposite behaviour between two segments of observations of count rates
$141$ and $113$ \ctss~(2-16 keV); \nub~increased from 0.11 to 2.3 Hz.
Therefore, this means that like in the case of the millisecond pulsar
SAXJ1808.4-3658 (Wijnands \& Van der Klis 1998), it appears that
\nub~correlates with the source count rate (the latter most likely
tracking the mass accretion rate), down to a certain value, below
which \nub~and the source count rate anti-correlates.  For both
\slx~and the millisecond pulsar, this transition occurred at a
luminosity of $\sim 2-4 \times 10^{36}$ \ergs~(3-25 keV).

\ks~is characterized by weaker variability with an integrated power of
$\sim 3.0$\% (2-40 keV). Its PDS can be approximated by a PL of index
$\alpha=1.3$ on which a QPO centered at $7.6 \pm 0.7$ Hz is superposed
(FWHM = $3.1 \pm 2.0$ Hz for a gaussian fit). This QPO is very similar
to the so-called ``Horizontal branch'' QPO observed in Z sources. In
addition, at higher frequencies, for the segment of the observation
performed on October 28th (i.e.  at the lowest count rate $\sim 1050$
\ctss, 5-30 keV), we have detected at the $4.7\sigma$ level a HFQPO
centered at $\sim 1200\pm10$ Hz (FWHM=$45.0\pm20$ Hz,
RMS=$2.5\pm0.4$\%)\footnote{The 1200 Hz HFQPO observed is a lower
  limit of the Keplerian frequency at the innermost stable circular
  orbit (Kluzniak 1998, Miller et al. 1998). This translates into an
  upper limit on the mass of the NS: $ M_{NS} \la 1.83 \times
  (1+0.75j)$ \msol, where $j=cJ/GM^2$ is a dimensionless stellar
  angular momentum, and $0.1<j<0.3$ for a NS spin frequency of $\sim
  260$ Hz (Kluzniak 1998). This yields an upper limit of 2.25 \msol~
  for the NS in \ks.}. This is the highest HFQPO ever detected from
\ks. No HFQPOs were detected in the subsequent observations performed
on October 29th (1285 \ctss) and 30th (1355 \ctss).  We have derived
an upper limit of $\sim 2$\% on the RMS of any HFQPOs around 1000 Hz
for the October 29th and 30th observations, respectively.  Our October
28th detection follows the correlation between the frequency of the
HFQPO (\nuhfqpo) and the count rate, as well as the anticorrelation
between \nuhfqpo~and RMS presented in Wijnands \& Van der Klis
(1997).

\subsection{Spectral fitting}
Previous observations of similar NSs have shown that their broad band
spectra can be generally fitted by the sum of two components (Mitsuda
et al.  1984, White et al.  1988, Mitsuda et al.  1989).  The first
one is soft, contributes to the spectrum mainly below 10 keV, is
modeled by a multi-color disk blackbody or a single temperature
blackbody, and is interpreted as coming either from the accretion
disk, or the NS surface, or an optically thick boundary layer (e.g.
White et al.  1988, Mitsuda et al.  1984).  The second one is harder
and dominates the spectrum above 10 keV; it is often modeled by simple
PLs (below 20 keV).  However, when high energy coverage exists above
30 keV (e.g.  with \sax~and RXTE), comptonization models provide more
physical fits (e.g. Guainazzi et al.  1998, In't Zand et al.  1999).
The comptonization process is speculated to take place in a scattering
corona located somewhere in the system; around the NS (e.g. optically
thin boundary layer, spherical corona) or above the disk.  In some
cases, a relatively strong iron \kalpha~line (6.4 keV) is also
observed above the continuum (e.g. White et al.  1986).  In a few
cases (e.g.  4U1608-522, Yoshida et al.  1993, SAXJ1808.4-3658,
Gilfanov et al.  1998), such a line is accompanied by a broad
absorption like feature, interpreted as partial absorption or
reflection by a cold or weakly ionised medium of the intrinsic PL
component.  Hereafter we report on the results of the spectral
analysis of our observations, placed in the framework of these
previous results.

\subsubsection{\onee} 

\begin{table*}[!th]
\begin{center}
\begin{tabular}{lll}
\multicolumn{3}{c}{\onee} \\ 
\hline
 Parameter & C+BB+L & C+MCD+L \\ \hline
\kte & 28.1$^{+3.0}_{-3.0}$ & 25.6$^{+3.1}_{-2.0}$\\
\ktw & 1.1$^{+0.1}_{-0.1}$ & 1.6$^{+0.2}_{-0.2}$ \\
$\tau$ & 2.9$^{+0.2}_{-0.2}$ & 3.3$^{+0.3}_{-0.3}$ \\
\ktbb, \ktin & 0.6$^{+0.1}_{-0.1}$ & 1.2$^{+0.1}_{-0.2}$ \\
\rbb, \rincost & 10.1$^{+2.3}_{-1.1}$ & 2.8$^{+0.2}_{-0.3}$ \\
EqW & 21$_{-9}^{+10}$ & 27$_{-10}^{+10}$  \\
\chisq~(d.o.f) & 79.9 (80) & 85.7 (80) \\
\hline
\fsoft & 1.52& 1.58 \\
\fhard & 0.90  & 0.91 \\
\fbb~(\%) & 14.4  & 27.2\\
\hline
\end{tabular}
\label{tz2par}
    \caption{Best fit spectral results for \onee.  \nh~has been set to $1.2
      \times 10^{22}$ \nhv. The C+BB+L model is \comptt+Blackbody+a
      6.4 keV narrow line (\sigfe=0.1 keV), whereas C+MCD+L is
      \comptt+Multicolor Disk blackbody+a 6.4 keV narrow line.
      \kte~is the electron temperature in keV. \ktw~is the temperature
      of the seed photons in the \comptt~model (in keV).  $\tau$ is
      the optical depth of the spherical scattering cloud.  \ktbb~is
      the temperature of the blackbody, \ktin~is the inner disk color
      temperature derived from the MCD model.  \rbb~is the equivalent
      radius of the blackbody, whereas \rincost~is the projected inner
      disk radius (Mitsuda et al.  1984); they are both given in
      kilometers and scaled at the source distance (6.6 kpc).  EqW is
      the equivalent width of the iron line in eV. The 1-20 keV flux
      (\fsoft) and hard X-ray 20-200 keV flux (\fhard) are given in
      units of \dmn.  \fbb~is the contribution of soft component (BB
      or MCD) to the total luminosity computed in the 1-200 keV range.
      Errors in all the tables are quoted at the 90\% confidence level
      (\chisq=\chisq+2.7).}
\end{center}
\end{table*}

Preliminary results of the analysis of the \onee~spectral data can be
found in Olive et al. (1999).  In the present analysis, \nh~has been
set to the ASCA/\sax~value (i.e.  $1.2 \times 10^{22}$ \nhv).  Driven
by the recent \sax~results, we have found that the broad band spectrum
of \onee~can be adequately described by the sum of two components; a
soft and a hard Comptonized component well fitted by the {\comptt}
model in XSPEC (Titarchuk 1994, Guainazzi et al. 1998, see also Barret
et al. 1999).  The soft component is well described by a BB
(\chisq=88.1, 81 d.o.f). A MCD could fit it as well, although the
\chisq~is slightly larger (\chisq=92.3, 81 d.o.f).  For the
Comptonizing cloud, we have derived an electron temperature (\kte) of
$\sim 27-30$ keV and an optical depth of $\sim 2.8-2.9$ and $\sim 1.1$
assuming that it has a spherical or disk-like geometry respectively.
In both cases, a temperature of $\sim 1.0$ keV for the seed photons
was derived.  These parameters, which are listed in Table 1, are
strikingly close to those found by Guainazzi et al.  (1998),
suggesting that in addition to its low variability in intensity (see
Fig. 1), the source is also stable from the point of view of its
energy spectrum. For the soft component, the best fit parameters are
strikingly similar to those derived from the \sax~observations.

\begin{figure}[!th]
\centerline{\psfig{figure=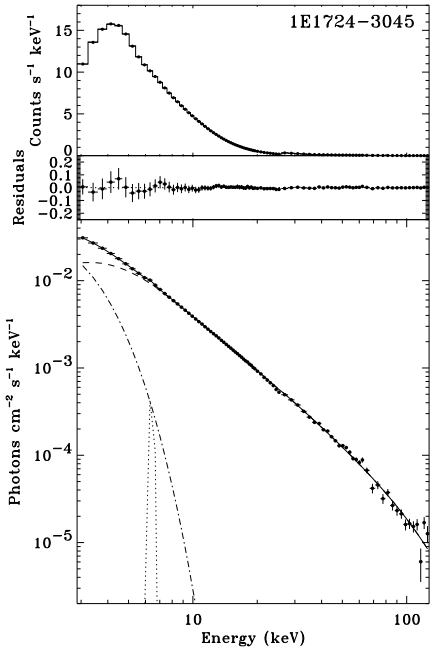,height=8.5cm}}
\caption{The data and folded model (top panel), the residuals
  (center panel) and the unfolded spectrum (bottom panel, Comptonized
  (\comptt) component is the dashed line, multi-color disk blackbody
  model is the dot-dashed line, the Gaussian line is the dotted line)
  of \onee.}
\label{sptz2}
\end{figure}

Independently of the model fitting the continuum, an excess around 6
keV seems to remain in the residuals.  Assuming that it results from
iron K fluorescence (6.4 keV), and that it is narrow (\sigfe=0.1 keV),
the \comptt+\bb+Line~model yields a \chisq~of 79.9 (80 d.o.f).  The
line equivalent width is 21 eV. In the remainder of this paper, we
will assess the significance of additional components in the fitted
model by means of the standard F-test.  What we call the F-test
probability will always refer to the probability of rejecting the
correct hypothesis, which states that, with the additional component
the fit is better.  We define \Fm~as $ (\chi^2_1-\chi^2_2) /
(\nu_1-\nu_2)/ (\chi_2^2/\nu_2)$ with $\chi^2_1 > \chi^2_2$ and $\nu_1
> \nu_2$ (where $\chi^2_{1,2}$ are the Chi-squared values of the two
fits, with $\nu_{1,2}$ number of degrees of freedom, \dof);
\Fm~follows a Snedecor's F distribution.  In the present case, we have
\Fm=8.2, and \pfm=0.005, which means that the addition of the line to
the \comptt+\bb~model is significant at the level of $\sim 99.5$\%.
It is worth pointing out that the significance of any features in the
spectrum depends strongly on the level of systematics added to the
data, and that there are some uncertainties in the estimate of these
systematics.  To illustrate this, if instead of using 0.5\% around 6
keV, one uses 1\%, as for example in Rothschild et al.  (1999), the
significance of the line would drop down to 85\%, and hence the line
detection would not be significant.  Leaving the line energy and its
width as free parameters does not improve the fit significantly.  The
presence of the line is also significant at the level of 98.5\% for
the \comptt+MCD model.  However, since no such a line was found in
ASCA and \sax~observations, we consider our detection very marginal.
Fig. \ref{sptz2} shows the data and folded model (\comptt+BB+L), the
residuals and the unfolded spectrum of \onee.

For comparison with SIGMA data (35-200 keV), we have fitted the HEXTE
data alone in the 25-150 keV range.  Although the fit is not
satisfactory due to the presence of a clear high energy cutoff ($\sim
70$ keV), a PL fit yields a photon index of 2.7 $\pm 0.1$, to be
compared with the time-averaged value observed by SIGMA ($3.0 \pm
0.3$, Goldwurm et al.  1993, 1994).  Fitting the 2.5-25 keV PCA
spectrum with a simple PL (note an acceptable fit) yields a photon
index of $\sim 2.0$.  We therefore conclude that the steepness of the
hard tail observed by SIGMA was artificial, and due to the presence of
a high energy cutoff around 60-70 keV. We have also fitted the
continuum with the relativistic Comptonization model developed by
Poutanen \& Svensson (1996) (\compps~in XSPEC).  For the Comptonized
tail, this yields best fit parameters of 35 keV for \kte~and 2.1 for
$\tau$ (spherical geometry assumed).

From this observation, we derive \lx$=8.1 \times 10^{36}$ \ergs, and
\lhx$=4.8 \times 10^{36}$ \ergs~(d=6.6 kpc). \lx~is consistent with
the time-averaged value derived from the ASM light curve.

\subsubsection{\gs} 
\begin{table*}[!th]
\begin{center}
\begin{tabular}{lll}
\multicolumn{3}{c}{\gs} \\ \hline
 Parameter & CPL+L & CPL+R+L \\ \hline
\alphapl & 1.70$_{-0.02}^{+0.01}$ & 1.72$_{-0.01}^{+0.01}$ \\
\ecutoff & 98.7$_{-7.0}^{+8.0}$ & 90.0$_{-7.0}^{+8.0}$ \\
\relref & \ldots &  0.15$_{-0.04}^{+0.03}$ \\
Line Energy & 6.4 (fixed) & 6.1$_{-0.2}^{+0.3}$ \\
\sigfe & 0.1 (fixed) & 0.48$_{-0.23}^{+0.25}$\\
EqW & 36$_{-10}^{+8}$& 50$_{-13}^{+17}$ \\ 
\chisq~(d.o.f) & 149.4 (86) & 85.2 (83) \\
\hline
\fsoft & 1.43 & 1.44\\
\fhard & 1.17  & 1.16 \\
\hline
\end{tabular}
    \caption{Best fit spectral results for \gs.  \nh~has been set to $0.5
      \times 10^{22}$ \nhv. The CPL+L model is the sum of a Cutoff
      Power Law (\alphapl~is the photon index, \ecutoff~is the cutoff
      energy in keV), plus a narrow 6.4 keV line (\sigfe=0.1 keV).
      The CPL+L+R is the same but with the reflected component added
      (the CPL is substituted by the \pexrav~model in XSPEC, Magdziarz
      and Zdziarski 1995), and the line energy and width (\sigfe) left
      as free parameters of the fit.  \relref~is the reflection
      scaling factor normalized to the CPL. It should be equal to 1
      for an isotropic source above an infinite flat disk. Same units
      for \fsoft~and \fhard~as in Table 1.}
\end{center}
\label{gsbfit}
\end{table*}

\begin{figure}[!th]
  \centerline{\psfig{figure=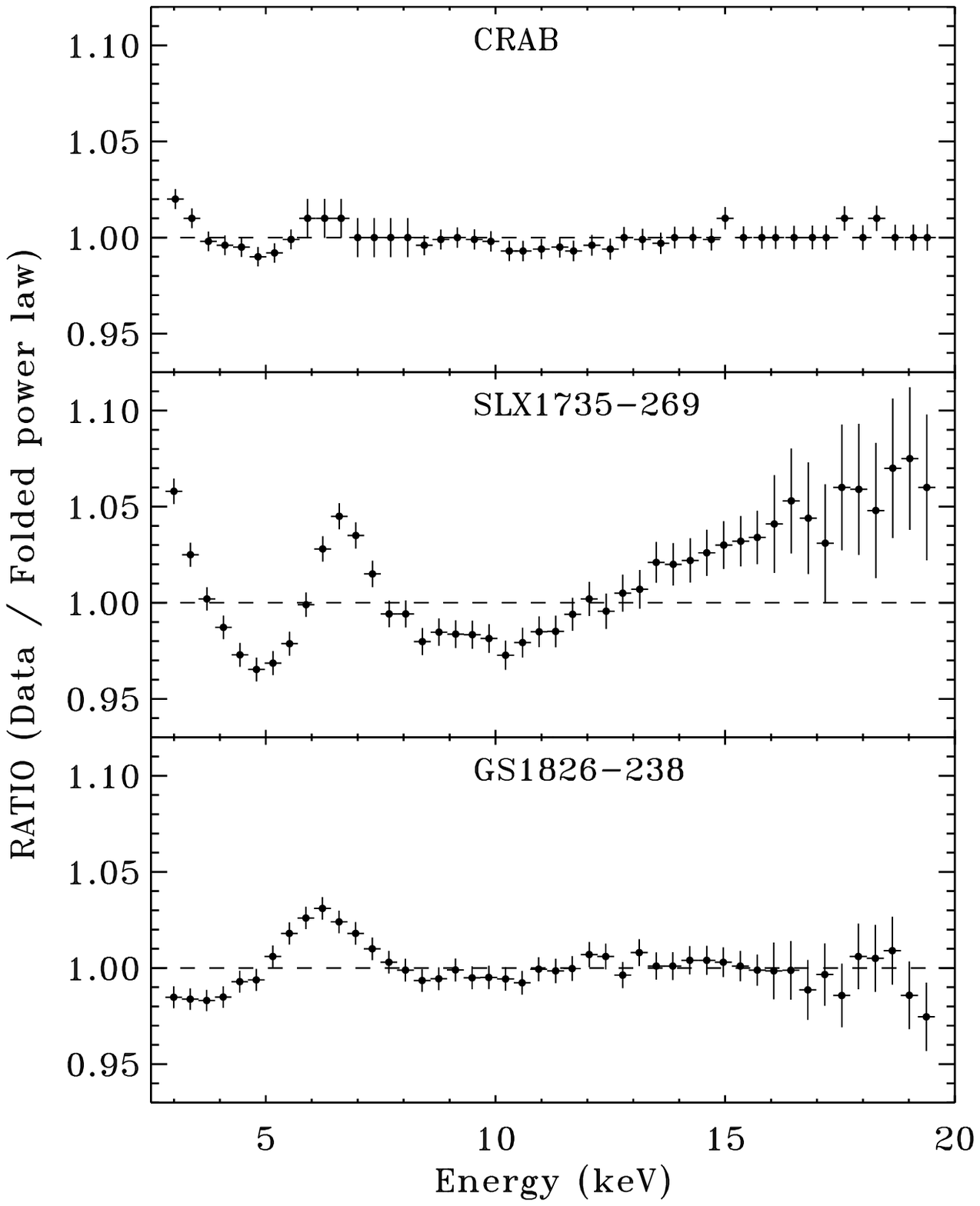,height=10.0cm}}
  \vspace*{-1cm}\caption{The ratio between the data and a folded power
    law model for the Crab, \slx~and \gs. The response matrice used in
    the Crab analysis has been obtained using the same procedure as
    for the other sources. For \gs~and \slx, this plot indicates the
    presence of a strong feature around 6-7 keV.}
\label{gs_ratio}
\end{figure}

For \gs, we have again set \nh~to the value observed by ROSAT and
\sax~($0.5 \times 10^{22}$ \nhv).  To illustrate the complexity of the
spectral shape, in Fig.  \ref{gs_ratio}, we show the ratio between the
PCA data and a PL model folded through the PCA response matrix.  There
is a clear excess around 6.0 keV.  First, of the single component
models, the CPL is the one that provides the best fit.  This yields a
cutoff energy at $\sim 95$ keV and a photon index of 1.7 (\chisq=195.0
for 87 d.o.f).  A fit using a comptonization model (e.g.  \compst~in
XSPEC) yields \kte~of 20 keV and an optical depth of 4.6.  Adding a
narrow 6.4 keV line (\sigfe=0.1 keV) to account for the feature of
Fig.  \ref{gs_ratio} improves the fit (\chisq=149.4 for 86 d.o.f) but
still, the residuals show systematic deviations that indicate that
this model is not the right description of our data.

Since a reflected component had been found in a combined analysis of
the OSSE and GINGA data (Strickman et al.  1996), we have tested for
the presence of such a component by adding an absorption edge at 7.1
keV. This leads to a significant decrease of \chisq~(\chisq=100.5 for
85 d.o.f, \Fm=41.3, \pfm$ \ga 7.9 \times 10^{-9}$).  Thus we have
substituted the CPL model by the XSPEC \pexrav~model \citep*{pexrav},
which is the sum of a CPL plus a Compton reflected component.  The
inclination angle to the source is unknown, but since \gs~is not a
dipper, we have assumed a standard value of $\theta = 60^\circ$, and
we leave the reflection scaling factor as a free parameter (we assumed
solar abundance for the reflecting material).  This yields a
reflection scaling factor of $\sim 0.20$, and a narrow 6.4 keV line
(\sigfe=0.1 keV) of EqW of 37 eV (\chisq=97.1, 85 \dof).  The
significance of the reflection component is very high (\Fm=45.4, \pfm
$ \ga 1.8 \times 10^{-9}$).  Leaving the line energy and its width as
free parameters of the model, we get \chisq=85.2 (83 \dof) with the
line parameters: \sigfe=0.48 keV, centroid energy at
$6.1^{+0.3}_{-0.1}$ keV (90\% confidence level, still consistent with
a fluorescent iron K$\alpha$ line), and an EqW of 50 eV. This decrease
of \chisq~is significant at the level of 99.7\% (\Fm=6.0, \pfm=0.003),
we therefore conclude that this is the best fit to our data.  As a
further step, we have tried a reflection model including ionisation:
\pexriv~model in XSPEC \citep*{pexrav}.  This model fits the data as
well as the \pexrav~ model (\chisq=95.5, 84 \dof), and yields a disk
ionisation parameter consistent with 0.  Therefore, with both models,
our data are consistent with reflection from a cool neutral medium.

\begin{figure}[!th]
  \centerline{\psfig{figure=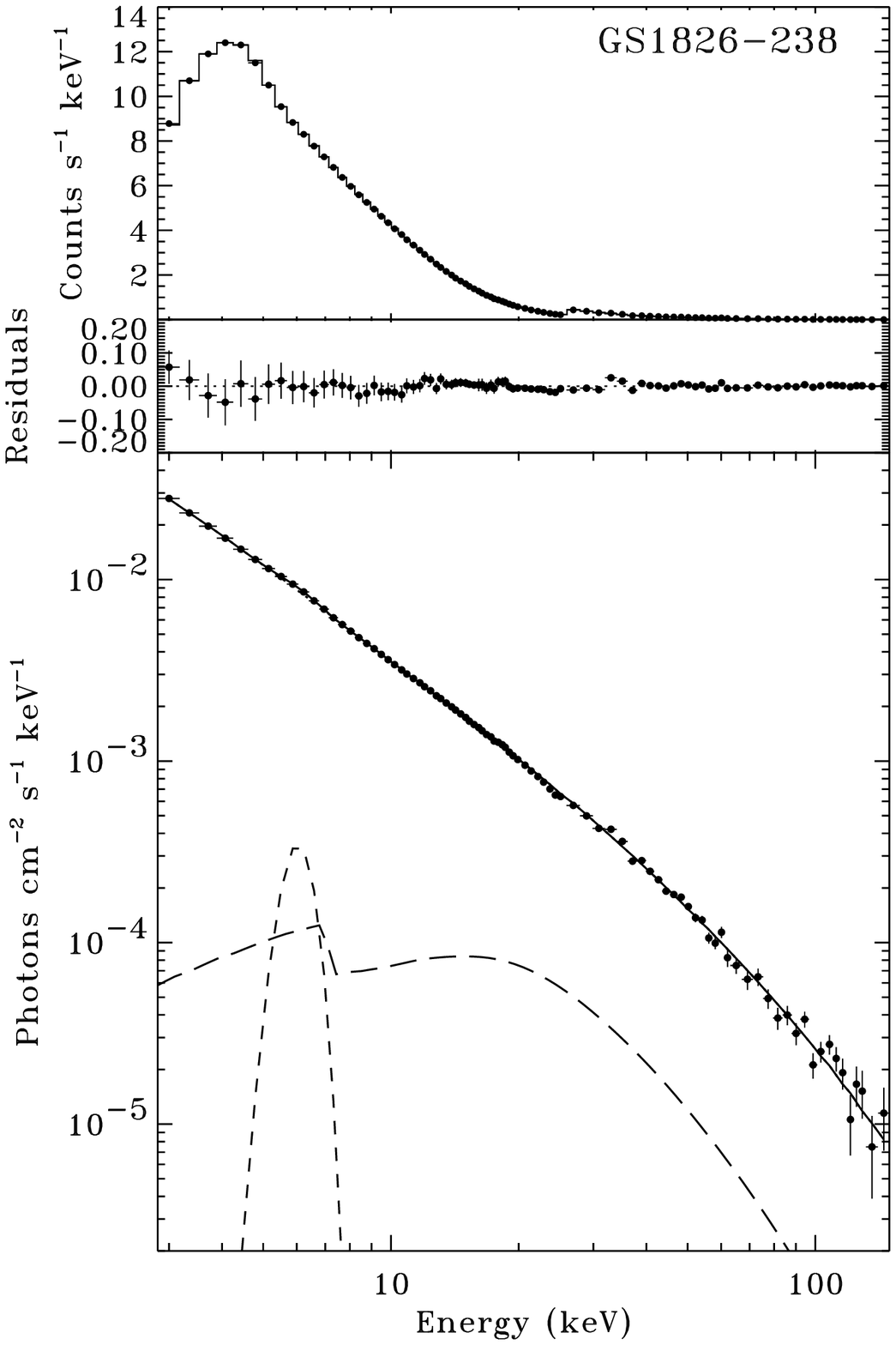,height=8.5cm}}
\caption{The data and folded model (top panel), the residuals
  (center panel) and the unfolded spectrum (bottom panel) of \gs. The
  iron line at 6.4 keV is represented by the dashed line, the
  reflection component by the long dashed line.}
\label{spgs}
\end{figure}

\begin{table*}[!t]
\begin{center}
\begin{tabular}{lll}
\multicolumn{3}{c}{\slx} \\ \hline
Parameter & CPL+MCD+L   & CPL+MCD+R+L  \\ \hline
\alphapl & 2.06$_{-0.01}^{+0.01}$ & 2.09$_{-0.04}^{+0.03}$ \\
\ktin & 0.53$_{-0.07}^{+0.06}$ & 0.44$_{-0.08}^{+0.10}$ \\
\rincost & 9.4$_{-2.9}^{+5.4}$ & 17.0$_{-9.1}^{+16.6}$ \\
\relref & \ldots &  0.28$_{-0.12}^{+0.09}$ \\
$\xi_i$ & \ldots & 70.1$_{-70.1}^{+740.2}$ \\
EqW & 67$_{-157}^{+15}$ &  39$_{-14}^{+15}$ \\ 
\chisq~(d.o.f) & 45.2 (59) & 28.4 (57) \\
\hline
\fsoft & 0.72& 0.75 \\
\fhard & 0.39 & 0.30 \\
\fbb (\%) & 9.0 & 12.6 \\
\hline
\end{tabular}
    \caption{Best fit spectral results for \slx.  \nh~has been set to 
      $1.5 \times 10^{22}$ \nhv. The first model is the sum of CPL+MCD
      and a 6.4 keV narrow line.  The best fit model is the sum of a
      CPL and its Compton reflected component (R) allowing for
      ionisation (\pexriv~in XSPEC), a MCD, and a narrow 6.4 keV line
      (L) (\sigfe=0.1 keV).  The cutoff energy in \pexriv~has been set
      to 200 keV and frozen during the fit.  $\xi_i$ is the ionisation
      parameter in \pexriv.  The contribution for the Galactic diffuse
      component has been modeled by a Raymond-Smith model of 3.4 keV
      (normal abundance) and normalization factor of $1.5 \times
      10^{-2}$ in XSPEC units.  See Table 1 for definition of \fbb~and
      units of \fsoft~and \fhard.}
\end{center}
\end{table*}

We have also reanalyzed the GINGA spectrum of \gs~used in Strickman et
al.  (1996) and Zdziarski et al.  (1999).  The reflection component is
highly significant, and consistent with coming also from a neutral
medium also.  A fit with a simple power law and a 6.4 keV narrow line
yields \chisq=39.6 for 31 \dof, while adding the reflection component
yields \chisq=16.1 for 29 \dof~(\Fm=21.2, \pfm=$2.1 \times 10^{-6}$).
The best fit parameters are \alphapl=$1.88\pm0.05$, and
\relref=$0.8\pm0.3$, and an equivalent width for the iron line of 15
eV\footnote{The evidence for a reflection component is very strong in
  both the RXTE and GINGA data.  It is therefore quite puzzling that
  no such component has been reported so far from the two
  \sax~observations performed, while \gs~was at a similar intensity
  level ($7.7\times 10^{-10}$ \ergscm~for RXTE, as opposed to
  $5.5\times 10^{-10}$ \ergscm~for \sax~(2-10 keV), Del Sordo et al.
  1999, In't Zand et al.  1999).  We are currently investigating this
  issue, searching for instrumental effects, sensitivity limits, and
  systematic errors in the fitting the continuum shape.}.

There is no soft component in the source spectrum.  With the \pexrav~
model, we have set a 90\% confidence limit of about 1\% on the
fraction of 1-200 keV luminosity in the soft component modeled by a
1.5 keV \bb~component.  The results of the best fit using the
CPL+Line+Reflection~model are given in Table 2. The data and folded
model, the residuals and the unfolded spectrum of \gs~are shown in
Fig. 9.

For comparison with hard X-ray observations of similar systems, we
have fitted the HEXTE spectrum with a PL. Although the fit is poor due
to the presence of a clear cutoff in the spectrum, one gets a photon
index of $ \sim 2.3 \pm 0.2$, similar to the value derived for \onee.
Fitting the hard tail with the \compps~model yields \kte=41 keV and
$\tau=1.9$.

For this observation, one derives \lx$ =8.4 \times 10^{36}$ \ergs, and
\lhx$ =6.9 \times 10^{36}$ \ergs~(d=7 kpc). \lx~is very close to the
time-averaged value derived from the ASM light curve.

\subsubsection{\slx} 

A preliminary report of the spectral analysis of \slx~can be found in
Skinner et al.  (1999).  As in the cases of \gs~and \onee, \slx~was
observed in a hard state.  However, it was a factor of two fainter
than the other two sources, so that the statistical quality of the
HEXTE spectrum is not as good.  For that reason, we considered the
HEXTE data only up to 50 keV. In our fits we fix \nh~ at $1.5\times
10^{22}$ \nhv, found in David et al.  (1997), as the ASCA data are
more sensitive to this parameter.  We have first fitted the PCA+HEXTE
spectrum of \slx~with a simple PL.  The fit is not good (\redchisq$\ga
6.5$) but shows clear evidence for a broad feature around 6.4 keV (see
Fig.  \ref{gs_ratio}).  As no iron line was seen in the ASCA data, and
because of the relative faintness of the source, we have first
considered the possibility that the line might not be from the source
but might represent the flux within the PCA 1\degree\ field of view
from the galactic bulge diffuse emission.  We have therefore included
in our model a contribution from the spectrum given by Valinia \&
Marshall (1998) for their ``Region 2'' which includes this part of the
sky.  Refitting the data with this component affects the shape of the
residuals at low energies ($\la$ 5 keV); they now clearly show the
presence of a soft component and reveal that the broad feature between
6-7 keV cannot all be accounted for by the diffuse emission, meaning
that there remains a need for a significant contribution from the
source to the line.  Such a line was also needed in the data analyzed
by Wijnands \& Van der Klis (1999).  The soft component was fitted
with a \dbb~(although the \bb~model fits the data equally well),
whereas the line contribution from \slx~was modeled by a gaussian
centered at 6.4 keV. This leads to a \chisq~of 45.2\footnote{For \slx,
  the \chisq~associated with the best fits are lower than 1 (see Table
  3).  This makes naturally questionable the results of our F-test.
  Similarly, the errors computed on the best fit parameters should not
  be considered as true 90\% uncertainties (these errors are computed
  under the assumption that the errors on the data are Poissonian).
  The low \chisq~indicates certainly that for \slx, which is the
  faintest of the 4 sources, the systematics assumed are too large.
  However, for consistency in the analysis, we have chosen to set them
  to the values used for the other sources.} (59 d.o.f), and thus the
addition of the line and soft component is statistically significant
based on our F-test.  The EqW found (67 eV if the line is narrow,
\sigfe=0.1 keV) is well below the upper limits of 150 eV (\sigfe=0.1
keV) derived by David et al.  (1997) using ASCA SIS data.

\begin{figure}[!th]
\centerline{\psfig{figure=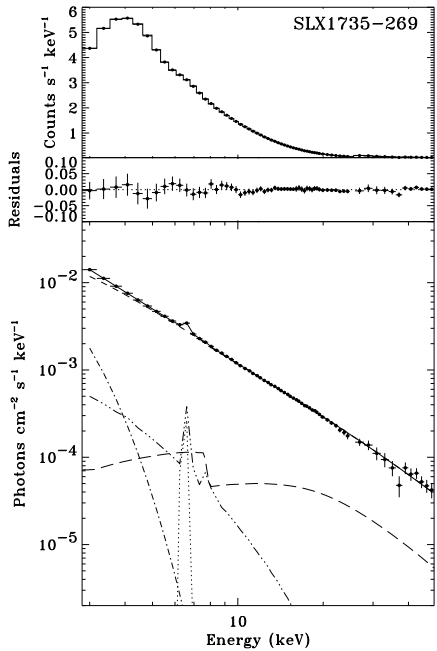,height=8.5cm}}
\caption{The data and folded model (top panel), the residuals
  (center panel) and the unfolded spectrum (bottom panel) of \slx. The
  power law component is represented by the dashed line, the
  multi-color disk blackbody model by the dot-dashed line, the iron
  6.4 keV line by the dotted line and the reflection component by the
  long-dashed line. The Raymond-Smith thermal spectrum corresponding
  to the galactic diffuse emission around \slx~is also indicated as a
  dot-dot-dot dashed line.}
\label{spslx}
\end{figure}
 
However, by looking at the residuals above $\sim 7$ keV the data show
systematic deviations with an edge-like shape.  An edge around these
energies is suggestive of the presence of a reflection component. We
have therefore first substituted the PL model with the \pexrav~model
in XSPEC (assuming no energy cutoff in the model, and a 6.4 keV line
with a \sigfe=0.1 keV, an inclination angle of 60\deg).  This leads to
\chisq=31.4 (58 \dof), indicating that the presence of a reflected
component is significant at a level greater than 99.99\% (\Fm=23.1,
\pfm$=1.1~10^{-5}$). The solid angle inferred is $\sim 0.26$ and the
line EqW is 47 eV.  Whether or not the reflector is ionised has been
tested by using the \pexriv~model.  The statistical quality of the
data does not allow us to determine the line energy simultaneously
with the Compton reflected component.  We have assumed the line at 6.5
keV (expected from a moderately ionised medium). This yields a
\chisq=28.4 (57 \dof), an ionisation parameter, albeit poorly
constrained, of $\xi_i \sim 70_{-70}^{+740}$, a solid angle of $\sim
0.28$ and a line EqW of 39 eV (see Table 3).  The improvement of
\chisq~from the \pexrav~to the \pexriv~model is significant at the
level of 98.3\% (\Fm=8.8, \pfm$=4 \times 10^{-3}$). It is worth
pointing out that leaving out the line at 6.4 keV increases the
significance of the ionisation of the reflector, but then the model
becomes unrealistic, as the line is not expected to be at 6.4 keV for
the ionisation parameters derived from the fit ($\xi_i \ga 70$). To
conclude, one can say that our data for \slx~are consistent with
reflection from a neutral or moderately ionised medium.

In the absence of any cutoffs in the useful energy range of the
present measurements, comptonization models cannot really be used.  We
note however that a Comptonization model for which the electron
temperature would be $\sim 30$ keV (as seen in \onee) could fit the
continuum as well as the PL.  This would yield an optical depth of
$\sim 3$ for the scattering cloud.  Similar results were obtained by
combining simultaneous ASCA and SIGMA data (Goldwurm et al.  1995).
The best fit parameters are listed in Table 3 while the data and
folded model, the residuals and the unfolded spectrum are shown in
Fig.  \ref{spslx}.

For this observation, we derive \lx$=6.0 \times 10^{36}$ \ergs, and
\lhx$=3.5 \times 10^{36}$ \ergs~(d=8.5 kpc). \lx~is about a factor of
two lower than the time-averaged value derived from the ASM light
curve.

\subsubsection{\ks} 
\begin{table*}[!th]
\begin{center}
\begin{tabular}{lllllll}
\multicolumn{7}{c}{\ks} \\ \hline
& \multicolumn{2}{c}{28/10} & \multicolumn{2}{c}{29/10}&
\multicolumn{2}{c}{30/10} \\
\hline
 & \multicolumn{1}{c}{C+L+MCD} &  \multicolumn{1}{c}{C+L+BB} &  \multicolumn{1}{c}{C+L+MCD} &  \multicolumn{1}{c}{C+L+BB} &  \multicolumn{1}{c}{C+L+MCD} &  \multicolumn{1}{c}{C+L+BB} \\
\hline
\kte & 2.8$_{-0.1}^{+0.2}$ & 2.8$_{-0.1}^{+0.2}$ & 
2.7$_{-0.1}^{+0.1}$ & 2.6$_{-0.1}^{+0.1}$ & 
2.6$_{-0.1}^{+0.1}$ & 2.6$_{-0.1}^{+0.1}$\\
$\tau$ & 10.6$_{-0.6}^{+0.4}$ & 10.4$_{-0.3}^{+0.2}$ &
12.1$_{-0.7}^{+0.5}$ & 11.8$_{-0.3}^{+0.2}$ & 
13.3$_{-0.6}^{+0.5}$ & 12.6$_{-0.2}^{+0.5}$\\
\sigfe & 0.9$_{-0.2}^{+0.1}$ & 0.9$_{-0.2}^{+0.1}$  &
0.8$_{-0.1}^{+0.2}$ & 0.8$_{-0.2}^{+0.2}$ & 
0.8$_{-0.2}^{+0.2}$ & 0.7$_{-0.2}^{+0.3}$ \\
\ktin~or \ktbb & $1.8_{-0.2}^{+0.2}$ &1.3$_{-0.1}^{+0.1}$ &
1.7$_{-0.2}^{+0.2}$ & 1.2$_{-0.1}^{+0.1}$ & 
1.3$_{-0.1}^{+0.2}$ & 1.0$_{-0.1}^{+0.2}$ \\
\rincost~or \rbb & 2.2$_{-0.3}^{+0.5}$ & 3.8$_{-0.4}^{+0.5}$ &
2.5$_{-0.5}^{+0.8}$ & 3.8$_{-0.2}^{+1.2}$  & 
3.9$_{-1.5}^{+1.9}$ & 4.7$_{-1.4}^{+4.4}$\\
EqW & 154$_{-26}^{+29}$ & 136$_{-29}^{+32}$ & 
118$_{-25}^{+30}$ & 106$_{-27}^{+32}$ & 
94$_{-26}^{+35}$ & 86$_{-28}^{+37}$ \\
\chisq~(d.o.f) & 50.6 (51) & 48.6 (51) & 
45.7 (51) & 43.9 (51) & 
51.4 (51) & 51.2 (51)  \\
\hline
\fsoft & 8.4 & 8.7 & 9.4 & 9.6 & 9.5 & 9.6 \\
\fhard & 0.07 & 0.07 & 0.08 & 0.08 & 0.09 & 0.09 \\
\fbb (\%) & 13.5 & 6.3 & 13.0 & 5.9 & 9.8 & 3.2 \\
\hline
\end{tabular}
    \caption{Best fit spectral results for \ks~for three segments of
      the observation (Oct 28th, 29th, 30th).  \nh~has been set to
      $1.3 \times 10^{22}$ \nhv. The best fit model is the sum of a
      \compst~model (C), a MCD or a BB, and a 6.4 keV gaussian line of
      fitted width (\sigfe). \rincost~and \rbb~are scaled at 8.8 kpc.
      See Table 1 for definition of \fbb~and units of \fsoft~and
      \fhard.}
\end{center}
\end{table*}

\ks~was observed in a high state, and its spectrum is clearly much
softer than the other three sources examined before.  We have assumed
the \nh~measured by ROSAT and ASCA (i.e. $ 1.2 \times 10^{22}$ \nhv).
As the source intensity increased smoothly during our 3-day
observation, and its intensity variations are accompanied with
spectral variations (see Fig. 5), as said above we have made a PCA
spectrum for each of the three days. These PCA spectra were fitted
simultaneously with a single HEXTE spectrum averaged over the whole
observation to increase the statistics at high energies.

We first analyzed the October 28th spectrum.  As there is a clear
cutoff in the spectrum around 10 keV, we have first fitted the
continuum with a simple comptonization model (\compst).  This model
alone is clearly rejected (\chisq $\ga$ 977 for 55 d.o.f).  This is
mainly due to the presence of a broad feature again centered around
6.4 keV. We have therefore added an iron K fluorescence line (6.4 keV)
and fitted its width and intensity.  This improves the fit
significantly (\chisq $ = 156.2$ for 53 d.o.f).  However, looking at
the residuals indicates that the low energy part of the spectrum is
not well accounted for.  As in similar systems, a soft component
(either a blackbody or a disk blackbody) is usually observed (e.g.
White et al.  1988), we have included such a component in the fit.
This again improves the fit (\chisq=48.6 for 51 \dof, \Fm=56.5,
\pfm=$1.2 \times 10^{-13}$) for addition of the BB model.  We have
tried to fit the ``hard'' component with a MCD model (Mitsuda et al.
1984), instead of using \compst, but the fit is rejected, as the
reduced \chisq~exceeds 4. On the other hand, the soft component could
equally well be fitted with a MCD. For both models, the equivalent
radius of the BB and the projected inner disk radius of the MCD are
small; \rbb $\la$ 5 km and \rincost $\la$ 4 km respectively, smaller
than the NS radius, or the expected inner disk radius around a NS
(especially if the NS is as massive as 2\msol, see above). This does
not mean however that the models can be ruled out, because for
instance only a small fraction of the NS surface could be involved in
the emission, or alternatively some fraction of the disk flux could be
intercepted and scattered up in a corona.

We have also tried the \comptt~model for the Comptonized component.
However, the temperature for the seed photons (\ktw) is typically
0.2-0.3 keV, and is therefore unreliable since the peak of the Wien
law, at 3\ktw, is also below 2.5 keV (the peak is needed for \ktw~to
be determined reliably).  For the comptonizing cloud, \kte~are
consistent between \comptt~and \compst, whereas $\tau$ is a factor of
$\sim 2$ lower for \comptt~than for \compst~(spherical geometry
assumed).

The broadening of the line (due to Comptonization?) is significant.
Assuming the line is narrow (\sigfe=0.1 keV), we obtain a larger
\chisq~(87.1 for 52 d.o.f).  The hypothesis that the line is narrow is
therefore rejected at more than 99.99\% confidence level (\Fm=40.4,
\pfm=$5.6 \times 10^{-8}$).  If one leaves the energy of the line as a
free parameter, its value tends to move towards lower energies ($\sim
6.2$ keV, \chisq=46.1 for 50 d.o.f).  However, this shift of the line
energy is not significant (\Fm=2.1, \pfm=0.14).

\begin{figure}[!th]
\centerline{\psfig{figure=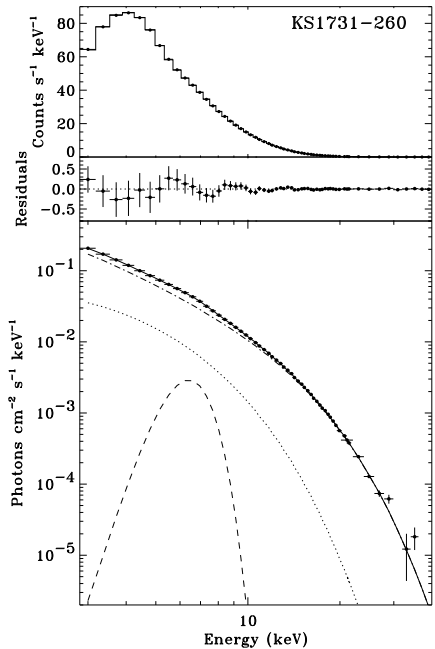,height=8.5cm}}
\caption{The data and folded model (top panel), the residuals
  (center panel) and the unfolded spectrum (bottom panel) of \ks~for
  the October 28th observation. The contribution from the
  \compst~component is shown by a dot-dashed line, the blackbody is
  the dotted line, whereas the broad iron line is shown with a dashed
  line.}
\label{spks}
\end{figure}

If an iron line of EqW of $\sim 100-150$ eV is present, one might
expect to detect also a Compton reflection component.  First adding an
edge at 7.1 keV (cool iron) never improved the fit significantly
(significance always lower than 95\%, the line energy and width were
kept fixed).  Similarly, we tried to fit the data with the
\pexrav~model, as a substitute to the \compst~model.  This never
improved the fit either.  The non detection of such a component might
be simply related to the curvey shape of the spectrum (in particular
the continuum is not a CPL as assumed in \pexrav); reflection is
easier to observe when the continuum is PL-like (Note that if there is
indeed a reflection component; this will affect the fitted parameters
for the continuum and hence the strength of the iron line).

The same model is also the best fit of the October 29th and 30th
spectra.  In Table 4, we list the best fit parameters for the
\compst+\bb+Line and \compst+\dbb+Line models.  The data and folded
model, the residuals and unfolded spectrum \ks~for October 28th are
shown in Fig.  \ref{spks}.

For \ks, \lx~was $\sim 8.0 \times 10^{37}$\ergs~at the beginning of
the observation and reached $\sim 9.1 \times 10^{37}$\ergs~at the end
(d=8.8 kpc). Such X-ray luminosity is slightly larger than the time
averaged ASM value. Throughout the observation, \lhx~contributed just
at the level of$\sim 1$\% to the 1-200 keV luminosity of the source.
\section{Discussion}
We have here reported RXTE PCA+HEXTE timing and spectral observations
of four type I X-ray bursters (hence NSs).  In the remainder, first we
discuss their timing properties, then their broad band spectral
properties, and finally address issues concerning possible differences
between BHs and NSs.

\subsection{Timing properties}
\onee, \slx~and GS1826-238 were in a so-called low state (LS) whereas
\ks~was in a high state (HS).  The LS sources all have very similar
PDS, characterized by High Frequency Noise (HFN, or flat top noise,
Van der Klis 1994) with large RMS amplitudes (30\%, 2-40 keV).
\onee~and \slx~ are characterized by the presence of a $\sim 1$ Hz
QPO, and a break (at \nub) at 0.1-0.2 Hz. For \gs, depending on the
modeling of the PDS, \nub~$\sim 0.02$ Hz or $\sim 0.2$ Hz, and
\nuqpo~$\sim 0.2$ or $\sim 1.0$ Hz.  KS1731-260 has a very different
PDS, displaying both Very Low Frequency Noise (VLFN) and a QPO at
$\sim 7.6$ Hz.  These properties are all consistent with those of
``atoll'' sources at varying luminosities; the 3 LS sources would be
in the ``island'' state, whereas KS1731-260 would be in the ``banana''
state, according to the terminology defined by Hasinger \& Van der
Klis (1989).
\subsubsection{Origin of the high frequency noise?}
The high frequency noise is seen only in sources displaying
significant hard X-ray emission (all sources except \ks), and hence
must be related to the existence of a hot scattering cloud. This noise
in the PDS is very similar in shape and normalization to that seen
from LS BHs, strongly suggesting that the same physical
mechanism is responsible for the spectrum and variability in these
systems. Since the spectral shape strongly suggests thermal Compton up
scattering, then the variability must have something to do with
varying either the seed photons, or optical depth, or the dissipation
in the hot region.

The overall PDS shape can be described by a superposition of randomly
occuring flares (or ``shots''), generally with some distribution of event
durations.  Such ``shot noise'' models can give a good description of
the PDS (Lochner et al.  1991 and references therein, Miyamoto et al.
1992, Nowak et al.  1999, Olive et al.  1998), but are only a
phenomenological, rather than a physical, description of the
variability.  To get a physical description requires associating the
``shots'' to changes in the comptonizing cloud.

The simplest variability to envisage is a change of the soft photon
input.  Thermal comptonization involves multiple scattering of these
seed photons on the hot electrons, so the spectrum at lower energies
responds first, followed after several scattering timescales by the
higher energy spectrum. Such models predict that there should be time
lags between the hard and soft energy bands, but that these lags
(which are simply a measure of the size of the scattering cloud)
should be constant irrespective of whether the input variability was a
short flare or a longer event. This directly conflicts with the
observed lags in the BHCs (Miyamoto et al. 1988) and the NSs (Ford et
al 1999), which clearly show longer lags for longer timescale
variability (Miyamoto et al. 1991).

A satisfactory explanation of the long timescale lags is difficult.
If the variability arises from varying the seed photons then the
length of the lag directly implies that the region is large.  This led
Kazanas, Hua and Titarchuk (1997) to develop the ``Extended Atmosphere
Comptonization Model'' (see also Hua, Kazanas \& Cui 1999).  This
assumes a source of white noise at the center of the scattering cloud
which has a density profile $n(r)\propto 1/r$ out to radii of a few
light seconds. Their inhomogeneous density distribution appears to
match the observed time lags, and source spectra, but the physical
situation is very hard to envisage. The majority of the gravitational
potential energy is close to the compact object, so how can this power
a hot corona whose size is many orders of magnitude larger ? Another
weakness of that model, is that it produces different PDS at different
energies, contrary to what is observed (e.g. Nowak et al., 1999, Olive
et al. 1998).

An alternative model which can fit the PDS and time lags but with a
small source size has been developed by Poutanen \& Fabian (1999).
Here they associate the ``shots'' with magnetic flares above a cold
accretion disk, and use the spectral evolution of the flares to
produce the long lags. The flare begins with electrons being heated
but the background disk seed photon density is high, so the early time
flare spectrum is soft. As the heating progresses, the energy
dissipated in the flare dominates that of the disk, so the spectrum
becomes hard. The drawback with such models is that they rely on
details of the magnetic dissipation, which are not well known.
Nonetheless, it is encouraging that at least under some circumstances
the lags and PDS can be matched by the envisaged small source.

\subsubsection{Origin of low frequency QPOs?}

The origin of the $\sim 1$ Hz QPO (at \nuqpo) in the LS sources, once
suspected to be a BH signature, is also unknown, and is not accounted
for by the above models.  Chen \& Taam (1994) have proposed that they
might arise from disk luminosity oscillations resulting from thermal
viscous instabilities developing within the inner disk region.
Vikhlinin et al.  (1994) have also suggested that they could result
from a weak interaction between the ``shots'' when the instability is
triggered in a region of stable energy supply.  To keep the energy
released constant on large timescales, the appearance of a strong shot
should affect the amplitude and the probability of occurence of
subsequent ``shots''.  Recently, developing a model specific for NSs
(i.e.  not accounting for the similarities between BHs and NSs),
Titarchuk \& Osherovich (1999) have associated the QPOs with radial
oscillations in a boundary layer.

It has been shown that in many sources there exists a strong
correlation between \nub~and \nuqpo~(Wijnands \& Van der Klis 1999b).
Our three LS sources follow this correlation (Olive \& Barret 1999).
This correlation suggests that the longest fluctuations (scaling as
\nub$^{-1}$) and the QPOs are related to the same unknown physical
mechanism or are produced in regions interacting with each other.
Since the same correlation is observed among BHs and Z-sources, this
mechanism cannot be related to the presence of a hard surface or even
a small magnetosphere.  In addition, in several systems (e.g.
4U1608-52, Yoshida et al.  1993, see however above for \slx), a strict
correlation between \nub~and the inferred mass accretion rate has been
observed.  For BHs, Esin et al.  (1998) have speculated that the rapid
variability that is associated with the hard X-ray emission could
originate from an ADAF, with the variability timescales determined by
the viscous or dynamical timescales in the ADAF. They have further
proposed that the 1-10 Hz QPOs could result from the interactions at
the boundary (the so-called transition radius in ADAF terminology)
between the ADAF and the outer cool accretion disk, in which case the
\nuqpo~would be some multiples of the Keplerian frequency at the
transition radius.  Hence, a shrinking of the inner disk radius
(possibly due to an increase in the mass accretion rate through the
disk) should result in an increase of the QPO frequency.  The
correlation between the position of the inner disk radius inferred
from the kilo-Hz QPO frequency and \nuqpo, observed for instance in
4U1728-34 (Ford et al.  1998), is consistent with the above picture.
Furthermore, if \nub~is somehow related to the size of the ADAF, the
observed correlation between \nub~and \nuqpo~would imply that the ADAF
contracts when the accretion rate increases (possibly due to an
increase of cooling flux from the disk).

\subsubsection{Spectral states and high frequency QPOs}

Finally, in the three sources displaying high frequency noise and a
hard X-ray tail, no HFQPOs (above 300 Hz) are detected (e.g \onee,
Barret et al.  1999b).  More generally, it seems that no HFQPOs are
seen when \nub~is low ($\la 6-7$ Hz; e.g. 4U1705-44, Ford et al.
1998).  The lack of HFQPOs might therefore be related to the presence
of a hot scattering corona.  Smearing of the HFQPO signal in such a
corona is a possible mechanism, which is however invoked at higher
accretion rates (i.e. for larger optical depth $\tau \ga 5$, cooler
corona, Brainerd \& Lamb, 1987).  This might indeed explain the
disappearance of the HFQPOs after it has reached saturation (i.e. when
the inner disk is at the last stable orbit, e.g. 4U1820-30, Bloser et
al.  1999), or more generally when the source enters the upper branch
of the ``banana'' state (Miller et al.  1998).  The anticorrelation
between the QPO amplitude and the QPO frequency observed in \ks~(see
above) is consistent with this picture (note that there is a weak
indication that the optical depth derived from the spectral fitting
increases when the source luminosity increases, see Table 4).  For the
LS sources, from the spectral analysis, we have derived an optical
depth of a few for the Comptonizing cloud.  Could that cloud, through
smearing, account for the lack of HFQPOs?  Using our upper limit, one
can set a lower limit on the size of the scattering region.  If the
radiation is scattered, the RMS amplitude at infinity ($A_{\infty}$)
of a luminosity oscillation with frequency $\nu$ and amplitude A$_{0}$
at the center of a spherical region of radius \rc~and optical depth
$\tau$ is $A_{\infty} \approx (2^{\frac{3}{2}}xe^{-x}+e^{-\tau})
A_{0}$, where $x=\sqrt{3 \pi \nu R_{C} \tau / c}$ ($\nu$ is the
frequency of the oscillation, Kylafis \& Phinney, 1989; Miller et al.,
1998).  For a beaming oscillation, the attenuation is stronger (due to
the fact that the scattering process tends to isotropize the photon
distribution), and a factor $2/(1+\tau)$ has to be multiplied to the
first term of the previous equation (Kylafis \& Phinney, 1989).  Let
us first consider $\nu=300$ Hz, from the previous expression, taking
$A_{\infty}=2.5$\% (our upper limit), assuming $R_C = 200$ km, one can
set an upper limit $\sim 4$ and $\sim 6$\% on $A_{0}$ for a luminosity
and beaming oscillations, respectively.  For a signal at 1000 Hz,
these upper limits become 12\% and 20\% respectively.  Alternatively,
if we assume $A_{0}=10$\%, for the signal to be attenuated down to our
upper limit, the size of the corona has to be larger than 100 km and
130 km for a beaming and luminosity 1000 Hz oscillation.  Clearly,
this indicates that with the low optical depth derived for the
comptonizing cloud, and for any plausible sizes of such a cloud, the
attenuation is not very strong, and if signal there is, it has to be
intrinsically weak.

Another possibility which may explain the lack of HFQPOs in LS sources
might be related to the position of the inner disk radius.  Obviously,
if the HFQPOs are generated in the disk, and if the inner edge of the
disk lies at large radius (see below), no signals at such high
frequencies should be seen.  In that respect, it is worth pointing out
that the bump (around 10 Hz) in the PDS of \onee~(Fig.\ref{4pds}) was
interpreted by Psaltis et al.  (1999) as a low frequency kilo-Hz QPO
(the same bump is seen in \gs~and \slx). Their conclusion was derived
from the fact that the 0.8 Hz QPO and the bump at 10 Hz fitted in the
global correlation observed over several frequency decades between
\nuqpo~and the frequency of the lower kilo-Hz QPOs (Psaltis et al.
1999).

\subsection{Spectral properties}
To first order, the broad band LS spectra of NSs can be approximated
with a PL followed by an exponential cutoff at 50-80 keV.  There is
also a weak soft component in 2 of the 3 LS sources, with temperature
of $\sim 0.5-1$ keV contributing less than $\sim 30$\% of the total
luminosity.  The latter quantities, when derived from the fit of PCA
data are subject to uncertainties because the fit starts at 2.5 keV,
because they depend on the \nh~values assumed, and because there are
some calibrations uncertainties of the PCA at low energies (E $\la$ 5
keV).  However, as for example in the case of \onee, the values
derived from our spectral fitting are consistent with the ones
obtained from observations that are more sensitive to such a soft
component (e.g. by \sax~or ASCA, Guainazzi et al.  1998, Barret et
al., 1999a).  Thermal Comptonization of these soft photons by hot
electrons then appears to be the most plausible emission mechanism for
these systems, and for a spherical scattering cloud, the derived
optical depths and electron temperatures are in the ranges of $2\la
\tau \la 4$ and $15\la$ \kte~$ \la 30$ keV. Thermal Comptonization
also dominates the spectral formation in \ks, but with a significantly
lower \kte~($\sim 3$ keV) and larger $\tau$ ($\sim 10$).  The main
difference between a source in the HS and a source in the LS is best
illustrated in Fig. 12, where a $\nu$F$\nu$ plot of the PCA/HEXTE
spectra of \ks~and \gs~are shown. Clearly for \ks, the bulk of the
energy is radiated below $\sim 10$ keV, whereas for \gs~the energy
spectrum is flatter and a large fraction ($\sim 50$\%) of the energy
is radiated in the hard X-ray band.

\begin{figure}[!t]
\vspace*{-1.0cm}\hspace*{-1.5cm}\psfig{figure=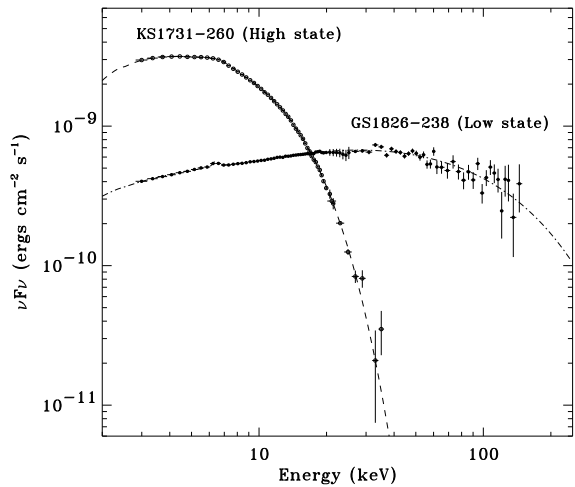,width=10cm}
\vspace*{-2.25cm}\caption{A $\nu$F$\nu$ plot of the spectra of \ks~(high state) and
  \gs~(low state). For \ks, the bulk of the energy is radiated below
  $\sim 7$ keV. For \gs, on the other hand, its spectrum is relatively
  flat, peaks around 40 keV and drops rapidly above 100 keV. No
  scaling for the different source distances has been performed.}
\label{nufnu}
\end{figure}

\subsubsection{Iron line emission and reflection in NS LMXBs}

There is strong evidence for iron line emission in 2 out of the 3 LS
sources examined here (the evidence is weaker for \onee). The derived
line energy is consistent with 6.4--6.5 keV, i.e. fluorescence from
neutral/moderately ionised iron.  There are three possible sites for
line emission in LMXBs, the NS surface itself, the accretion
disk, and accretion disk coronal wind.  Any line from the NS
would be strongly redshifted, down to $\sim 5$ keV, while the
accretion disk corona should be strongly ionised, producing iron lines
at 6.7 and 7.0 keV. Thus, the most likely origin for the observed 6.4
keV line is irradiation of the accretion disk by the X--ray source.
If so, then it should be accompanied by a Compton reflected spectrum.
We significantly detect a neutral reflected continuum in \gs, and a
neutral/moderately ionised reflected continuum in \slx.

The reflection albedo, the ratio between the reflected and incident
luminosities, is then an {\it observationally} determined quantity
from our data.  The reflection probability is determined by the
relative importance of electron scattering and photo--electric
absorption.  At low energies the X--ray photons tend to be
photo--electrically absorbed rather than scattered, so the reflection
probability is low, while at higher energies the opposite is true. The
photo--electric absorption probabilities are determined by both the
element abundances (more heavy elements mean more hard X--ray opacity
and so less reflection) and ionisation state (high ionisation states
mean less bound electrons, so less opacity and more reflection).
Beyond 10-20 keV, the photo--electric opacities become negligible in
comparison to the electron scattering probability, but at these
energies the electron scattering is not elastic. The incoming photon
can lose much of its energy to the electron as it scatters, leading to
a marked decrease in the reflected probability above 50--100 keV.
Thus, even high energy photons can dump most of their energy in the
disk. For hard spectra, such as those seen in \gs, the {\it maximum}
albedo, even with complete ionisation of the disk is $a_x\sim 0.75$,
while the {\it observed} albedo is much less, at $\sim 0.25$, because
the ionisation state of the reflecting material is low/moderate, so
most of the low energy incident flux is photo--electrically absorbed
rather than reflected.  For the HS spectra, such as \ks, the
reflection albedo can be much higher if the disk is completely
ionised, since the spectrum is much softer.  This is important because
the disk heating and hence optical/UV reprocessed X--ray flux is
determined by the non reflected flux, $(1-a_x)$.  Various attempts to
derive the albedo from the optical reprocessed flux (e.g. de De Jong,
van Paradijs \& Augusteijn 1996) give $a_x\sim 0.9$. This is not in
conflict with our values since most of the LMXBs considered were in
the high state, where much more of the spectrum can be reflected.
However, we caution that the optical determinations assume that the
disk shape is that given by Vrtilek et al. (1990), with height
$\propto r^{9/7}$. This was derived assuming that the disk is
isothermal with height (central temperature equal to surface
temperature), which is incorrect (e.g.  Dubus et al. 1999).

While the fraction of incident to reflected flux depends on the X--ray
albedo (which in turn is given by the elemental abundances and
ionisation state of the reflector and the spectral shape of the
incident spectrum), the total amount of reflection seen depends on the
solid angle subtended by the material, $\Omega/2\pi=$\relref, and its
inclination to the line of sight. For an assumed inclination angle of
$60^\circ$, we derived \relref$=0.15$ and $0.28$ in \gs~and \slx,
respectively, with an associated iron line EqW of 50 and $39$ eV.  The
line EqW and especially the reflected fraction are significantly lower
than the $\sim 130$ eV and \relref$=1$ for an isotropic X--ray source
above a flat, infinite disk (George \& Fabian 1991; Matt, Perola \&
Piro 1991).

\subsubsection{Accretion Geometry}
Both the broad band continuum spectrum, and its variability (e.g. van
der Klis 1994; Ford et al. 1999) and the properties of the reflected
spectrum (Zycki, Done \& Smith 1998, 1999) are very similar to that
seen in the LS BH systems, suggesting that the same physical
mechanisms operate in a similar accretion geometry.  If so, then the
hard emission cannot have anything to do with the magnetosphere or the
NS surface, and must instead be connected with the accretion flow.
Mechanisms suggested for the hard power law in BH systems which can
also work in NSs are either an accretion disk corona (magnetic
reconnection in flares above a disk), or a hot accretion flow.  The
lack of a strong reflection signature and weak soft emission is at
first sight incompatible with the magnetic flaring model.  A geometry
where the hard X--ray source is above an optically thick disk should
lead to \relref~$\sim 1$ rather than the \relref $\la 0.3$ observed
here.  In addition, of the hard X-ray flux, half will go up and escape
while half will go down and illuminate the cool material.  Of these,
10\% can be reflected as a hard X--ray component (if the material is
mostly neutral, and has solar abundances), but the remaining 90\% are
thermalised, and should emerge as a soft component.  The soft photon
luminosity would then be roughly half of that of the hard component.
However, it is only emitted into $2\pi$ solid angle because it is
optically thick, and since the hard component is emitted into $4\pi$;
the inferred luminosities of the soft and hard components should be
about equal (see Gierlinski et al.  1997).

The properties of the reflected component and line emission together
with the observation of time smearing or delays in the optical
counterparts of X-ray bursts testify of the presence of an outer disk
in those systems.  It would seem however that a significant fraction
of the disk cannot be seen, either in reflection, or by reprocessing
the hard X--ray radiation, or by emission.  So the question is: is
there then an inner disk at all, say within $\sim 50-100$ \rs~?  If
not, then the disk corona model can be ruled out.  However, the disk
reflection and reprocessing signatures could perhaps be masked by a
very highly ionised inner disk (Ross et al.  1998, but see Done \&
Zycki 1998), while the intrinsic disk emission is then cooler and less
intense if the power is mostly dissipated in the corona (Svensson \&
Zdziarski 1994).  Another possibility is that the hot plasma has a
bulk velocity $\beta c$ directed away from the disk (Beloborodov
1999).  The bulk motion might be due to the pressure of the reflected
radiation and/or plasma ejection from magnetic flares.  A mildly
relativistic escaping flow ($\beta\sim 0.3$) causes aberration which
reduces the downward irradiating flux which in turn reduces the
feedback of reflection and reprocessing.

Perhaps the most telling discrepancy in all these models is that there
should also be strong emission from a boundary layer where the disk
and NS interact. The boundary layer emission can be twice as large as
that from the accretion disk (whether standard disk or disk powered
corona) once relativistic corrections are included (Shakura \& Sunyaev
1986).  There is no strong soft emission component in the spectrum,
ruling out an optically thick boundary layer (unless it also
dissipates most of the accretion energy in an optically thin region
perhaps powered by magnetic reconnection). King \& Lasota (1987)
calculated that the boundary layer between a standard disk and the NS
can be optically thin only for mass accretion rates much lower than
those considered here (but see recent calculations by Inogamov \&
Sunyaev, 1999 and Popham \& Sunyaev 1999). Alternatively, if the disk
truncates at the last stable orbit (allowed only for certain NS
equations of state) then the boundary layer is between material freely
spiraling in rather than a disk. This can give an optically thin
boundary layer up to luminosities of a few percent of Eddington, as
required here (Kluzniak \& Wilson 1991). However, it seems unlikely
that an optically thin, hot boundary layer could be similar in
temperature and variability to disk emission produced by a completely
different process of magnetic flare. Removing the boundary layer by
ejecting the material just before it reaches the NS surface (e.g. by a
centrifugal magnetic field barrier or winds) is not a viable way
around these difficulties.  The observation of X--ray bursts from
these systems proves that the accreting material does accumulate on
the NS surface (e.g. \gs).

Such a series of requirements make the disk corona models seem rather
contrived, so we investigate the alternative model where the inner
disk is replaced by an X--ray emitting hot flow.  This would naturally
account for the weakness of reflection and reprocessing, and the
relative weakness of the intrinsic soft emission.  It also provides an
explanation for the strong correlation observed between the relative
strength of the reflection component (\refrel), and the photon index
of the intrinsic PL (\alphapl) in Seyfert galaxies and galactic BHs
(Zdziarski et al.  1999).  The same correlation also applies to NSs,
including those considered here; for \gs,
\refrel=0.15$_{-0.04}^{+0.03}$ and \alphapl=1.72$_{-0.01}^{+0.01}$,
whereas for \slx, the corresponding values are
\refrel=0.28$_{-0.12}^{+0.09}$ and \alphapl=2.09$_{-0.04}^{+0.03}$).
This correlation can be understood by considering a hot X-ray plasma
that fills a variable-radius hole in the inner disk; as the disk
radius decreases, its solid angle increases, thereby increasing both
the reflection and the flux of cool cooling photons. The stronger
cooling of the hot plasma then leads to a steepening of the PL slope
(Zycki, Done \& Smith 1998, Zdziarski et al. 1999)\footnote{
  Alternatively, this correlation can also be explained by the outflow
  model of Beloborodov (1999).  Increasing the semi-relativistic
  outflow velocity, $\beta$, leads to a decrease of both \alphapl~and
  \relref, and yields a correlation that seems to fit better the
  observations than the two phase disk model (Zdziarski et al.  1999).
  Another interesting feature of the ejection model, is that it can
  explain \relref~$\ga 1$ (as observed in some Seyferts) because for
  negative values of $\beta$, the coupling between the ejected plasma
  and the disk can be very strong.  As said above, however, the weak
  point of such a model is that it relies on the poorly known physics
  of the magnetic dissipation within the disk.}.

The above scenario, where the innermost radius of the accretion disk
moves as a function of spectral state, can also naturally explain the
strong correlation between spectral properties and the frequency of
the HFQPOs (e.g. Kaaret et al. 1998; Mendez et al. 1999, Bloser et al.
1999).  While there are several current models proposed for these high
frequency features in the power spectrum, they all require that the NS
spin beats with an inner disk frequency (see e.g. the review by Van
der Klis 1998). The easiest way to change the inner disk frequency is
to change the inner disk radius. This then naturally explains the lack
of HFQPOs in the LS NSs as being systems where the inner disk is
truncated at large radii, where the interaction with the NS surface is
small. As the disk moves inwards, the spectrum softens
(\alphapl~increases; e.g.  in 4U0614+09 and 4U1608-52, Kaaret et al.
1998, and the disk contribution increases; e.g. 4U0614+09, Ford et
al., 1997), the HFQPOs appear, and their frequencies increase (Piraino
et al., 1999).

How can we reconcile our spectral and timing observations in that
picture?  \gs~ would have an accretion disk with a large inner radius.
Its inner disk temperature is then too low for its contribution to be
detected with the PCA. The cooling is low, its spectrum is hard and
the amount of reflection is rather small.  Conversely, for \slx, the
disk extends closer to the irradiating source, the solid angle for
reflection is larger and its spectrum is softer.  Finally for \ks, the
disk gets even closer; kilo-Hz QPOs are produced and are seen before
disappearing as a response to a further increase of the accretion
rate.  The cooling is strong and leads to a quenching of the hard
X-ray emission; at the same time the iron line gets stronger (so
probably, does the reflection component).  The above picture, although
attractive, does not explain why \onee, which has spectral and timing
parameters more or less similar to \slx~does not display any
reflection component (yet it shows a weak iron line).  One possible
explanation could be that the inclination is larger for \onee~than for
\slx.

\subsubsection{Emission Mechanisms}

But what can replace the inner accretion disk ?  The recent
rediscovery of a stable, X--ray hot solution of the accretion flow
equations has caused much excitement.  The main assumption of these
advective solutions is that the gravitational energy released by
viscosity is gained mainly by the protons, and that these heat the
electrons only by Coulomb collisions. For low mass accretion rates the
flow is optically thin so that the Coulomb collision rate is very low.
The proton temperature is high, so the flow has a large scale height
(quasi--spherical).  The small amount of energy that is gained by the
electrons is radiated as cyclo/synchrotron emission, bremsstrahlung
and Comptonization of these seed photon distributions (Narayan \& Yi
1995). Pure advective models have a disk existing only at very large
radii, where its emission is negligible. The resulting spectrum is
then rather hard, since the only seed photons are the self--produced
ones in the optically thin flow, and the electron temperature is of
order $\sim 100$ keV (Narayan \& Yi 1995). As the mass accretion rate
increases, the flow becomes denser, so the Coulomb collisions are more
effective at transfering energy from the protons to the electrons, so
the radiative efficiency increases. But this process cannot continue
indefinitely: as the flow becomes optically thick, the Coulomb
collisions drain all the energy from the protons, and the flow
collapses into a standard, optically thick, geometrically thin
accretion disk (Esin et al. 1998).

However, there is a clear difference between advective models for BH
and NS systems. For BHs the energy advected with the protons in the
flow can be swept invisibly down into the black hole. Such a flow will
collapse at $L\sim 0.4 \alpha^2$ \Ledd~where $\alpha\sim 0.2$ (Yi et
al. 1996) is the $\alpha$ disk viscosity.  For NSs the advected energy
is released in a boundary layer as the flow hits the NS surface. If
this boundary layer is optically thick, then the increase in seed
photons for the Compton cooling will cause the advective flow to
collapse at Eddington scaled mass accretion rates lower (by at least a
factor 3) than for BHs (Narayan \& Yi 1995; Yi et al.  1996).  Yet the
spectral state transition takes place at roughly the same Eddington
fraction for NSs and BHs, namely at $\sim 0.05-0.1$ \Ledd~(see e.g.
Mitsuda et al. 1989 for the NS 4U1608-52, and for the BH Cyg X-1, Esin
et al.  1998, see also Esin et al. 1997), and no strong soft emission
from a boundary layer is seen.  

One way out of this impasse is if the boundary layer is optically
thin.  Explicit calculations for the boundary layer between an ADAF
and NS surface have not yet been done, though rather different
spherical flows have been shown to give optically thin boundary layer
emission up to fairly high luminosities (e.g. Zane et al. 1998).  Thus
an optically thin boundary layer between an advective flow and NS
seems plausible.  In this case, the boundary layer is an additional
heating source for the hot plasma in the advective flow, so it might
be expected to merge rather smoothly in both spectral and variability
properties with the rest of the hot, optically thin accretion flow. 
The extra photons from the boundary layer are then hard.  This lowers
the temperature of the advective flow slightly, but the critical mass
accretion rate which can be sustained in the flow is then very similar
to that calculated for the BH case (Narayan \& Yi 1995).  The one
caveat to this is that even if the boundary layer were optically thin
then the NS surface would intercept and thermalize some of the
boundary layer emission, again leading to an additional source of soft
seed photons.  If the NS surface is ionised and/or mainly made up of
hydrogen due to settling of heavy elements, then the reflection albedo
can be as high as 0.6.  Thus at least 25\% of the boundary layer
luminosity should emerge as a soft component.  We are currently
calculating the effect of such an optically thin boundary layer plus
its thermalized emission on the properties of the advective accretion
flow, to see whether it can allow the observed similarity in state
transition accretion rate between BHs and NSs (Done \& Barret 1999).

It has been recently suggested that mass loss via winds was a natural
consequence of ADAFs, and was such, that only a tiny fraction ($\ll
1$) of the gas supplied through the ADAFs was actually accreted onto
the central object (Blandford and Begelman 1999, Quataert \& Narayan
1999).  Such winds would provide an alternative explanation for the
dimness of quiescent BHs, attributed in the ADAF models, to low
radiative efficiency and the existence of event horizons in those
systems (Narayan, Garcia \& McClintock 1997; Menou et al. 1999).
However, in our picture, where the NS is surrounded by an ADAF, the
observation of X-ray bursts tells us that all the matter flowing
through the ADAF can be accreted onto the NS. So, unless the mass
transfer rate is very much larger than currently thought in LMXBs (see
Menou et al. 1999 for a recent discussion), the presence of ADAFs
around bursting NSs would strongly argue against the existence of
powerful winds from such accretion flows (i.e.  ADIOS in Blandford \&
Begelman 1999).

To summarize, the similarity of the BH and NS spectra and variability
in their LS strongly suggest that the same mechanisms are operating
i.e. that the hard X--ray emission is connected to the accretion flow
with the NS surface having little impact on it.  Although the latter
remains an unescapable part of the system, there is nothing which can
be firmly associated with the strong emission expected from the
surface or the boundary layer even though it is known that the
accretion material has to accumulate onto the NS surface because of
X--ray bursts.  This problem applies to both the disk corona and
advective flow models. The boundary layer emission then must either be
optically thin, or emitted at too low a temperature to be observed.
The latter seems unlikely given previous ASCA and \sax~observations,
hence we conclude that the boundary layer is optically thin, and hot
(see recent calculations by Popham \& Sunyaev, 1999, Inogamov \&
Sunyaev, 1999).  However, it seems difficult to imagine a situation
where the boundary layer emission could have similar spectral and
variability properties to magnetic flares above a disk. Hence we favor
models where the inner disk is replaced by an X--ray hot, optically
thin flow. The only known stable hot solutions to the accretion
equations are the advective flows.  Unfortunately ADAF solutions have
not been developed with self-consistency for the NS case, and it is
not yet known whether these can have an optically thin boundary layer.
If so then this emission could well merge smoothly with the hot
optically thin advective flow. We stress that the maximum mass
accretion rate at which an advective flow can be sustained holds out
the possibility of showing whether or not advective flows can really
be present.  Observations show that the hard/soft spectral transition
associated in these models with the collapse of the advective flow
occurs at roughly the same mass accretion rate in both NS and BH
systems. While an optically thick boundary layer produces a large
difference in this critical mass accretion rate, an optically thin
boundary layer does not.  However, even an optically thin boundary
layer produces some soft photon flux from reprocessing of the flux
illuminating the NS surface, which will lower this critical mass
accretion rate.  Further calculations are needed to see whether this
effect is small enough to match the observed spectral transitions in
NSs.

\subsection{Comparison between BHs and NSs}
Thanks to RXTE and \sax, the number of LS NSs observed simultaneously
in X-rays and hard X-rays is growing rapidly, so that reliable
comparisons between BHs and NSs can now be carried out.  Although, as
illustrated in this paper, the most recent data indicate that BHs and
NSs are very similar in many respects, especially in their low states,
it remains critical to search for observational criteria that could
distinguish these two types of accreting systems.

\subsubsection{Spectral shape differences?}  
Heindl \& Smith (1998) have pointed out that the index of the power
law part of the spectrum is significantly larger for NSs than for BHCs
(\alphapl$\ga 1.8$ for NSs versus $1.4 \la $\alphapl$ \la 1.6$ for
BHs, see however Churazov et al.  1997 for an opposite conclusion).
The data presented here are in general terms consistent with this
claim, but indicates that the separation is by no means large (the
index for \gs~is 1.7).  On the other hand, Heindl \& Smith's claim, is
inconsistent with several recent observations of NSs, as for example
those of SAXJ1748.9-2021 in NGC 6440 (In't Zand et al.  1999) or those
of the two dippers 4U1915-05 (Church et al.  1998) and XB1323-619
(Balucinska-Church et al., 1999).  In the first case, a fit by a
broken PL yields a photon index of $1.54 \pm 0.03$ and $2.13\pm0.04$
below and above the break at $18.1\pm 1.2$ keV (In't Zand et al.
1999).  For 4U1915-05, the broad band 0.2-200 keV non-dip
\sax~spectrum can be fitted by a blackbody plus a cutoff PL of index
$1.6\pm0.01$ (\ecutoff$=80.4\pm10$ keV).  Similarly, for XB1323-619,
Balucinska-Church et al.  (1999) have found that the non-dip spectrum
is a cutoff power law with \alphapl$=1.48\pm0.01$ and \ecutoff=44.1
keV. Based on these observations, we therefore conclude that the above
criterion is not valid.

It has also been proposed that, in the framework of thermal
Comptonization models, the electron temperature of the scattering
cloud (\kte) appeared to be systematically lower for NSs than for BHs;
\kte $\la 30$ keV versus \kte $\ga 50$ keV (Tavani \& Barret 1997,
Zdziarski et al., 1998, Churazov et al.  1997).  This naturally
reflects the standard picture that on average BH spectra are harder
than NS spectra, and has been tentatively explained by the additional
cooling provided by the NS surface, which may act as a thermostat
capable of limiting the maximum \kte~achievable in these systems
(Kluzniak 1993, Sunyaev and Titarchuk 1989).  The data presented in
this paper are certainly consistent with this criterion.  However, it
is worth noting that there are some speculative BHCs (e.g.
GRS1758-258) for which \kte~derived from the fitting of their hard
X-ray spectra with the Sunyaev \& Titarchuk (1980) Comptonization
model (\compst~in XSPEC) is below 50 keV (\kte$\sim$33 keV in Mandrou
et al.  1994 from a fit of SIGMA data alone).  However, as pointed out
by Zdziarski et al.  (1998), the latter temperature is probably a
gross underestimate, and should be reevaluated using broad band
spectra and more appropriate relativistic Comptonization models
including Compton reflection and relativistic effects (e.g. Poutanen
\& Svensson 1996).  If such an underestimate has indeed been observed
in Cyg X-1 (\kte~increased from 27 keV with \compst~up to $\sim 100$
keV with \compps; Gierlinski et al.  1997), we note that in the case
of \onee, \kte~derived with \comptt~is 30 keV, whereas it is only
slightly larger ($\sim 35$ keV) with the relativistic \compps~model
(the same is true for \gs).  Bearing this in mind, it remains that
this criterion should be considered seriously for sources showing
evidence for thermal cutoffs in their hard spectra.  More data should
tell us soon whether all BHCs fitted with relativistic Comptonization
models will indeed have \kte~$\ga$ 50 keV, whereas all NSs fitted with
the same models will have \kte~below the above value.

In any case, the above criterion applies when an energy cutoff is
observed.  However such cutoffs are not always present in NS hard
X-ray spectra.  The first SIGMA observation of \onee~revealed a non
attenuated hard power law (\alphapl=1.8) extending up to 200 keV
(Barret et al.  1991).  Aql X-1 was also observed by BATSE from 20 keV
up to $\sim 100$ keV with \alphapl~in the range 2.1-2.6 and no
evidence for a high energy cutoff (Harmon et al.  1996).  Finally, a
recent \sax~observation of 4U0614+09 has placed a lower bound on any
exponential cutoff of the power law (\alphapl=2.3) of 200
keV\footnote{Alternatively, the \sax~spectrum could be modeled as
  thermal Comptonization (\compps~model in XSPEC, Poutanen \&
  Svensson, 1996) for which a very high \kte~was derived (246$
  ^{+50}_{-30}$ keV, Piraino et al.  1999).  Such a value, which puts
  the cutoff energy above 700 keV, is itself well above the high
  energy threshold of the \sax/PDS (the last significant data point of
  the spectrum is at 180 keV), and therefore must be considered with
  cautions.  However, if this result is confirmed by observations
  performed at even higher energies, it would make the \kte~criterion
  discussed above definitely unvalid.}  (Piraino et al.  1999).  This
means that just as there are two classes of BHs based on their hard
X-ray spectra (Grove et al.  1998), there might also be two classes of
NSs.  Members of the first class would display hard X-ray spectra with
energy cutoffs, which would result from thermal Comptonization.
Members of the second class would have non-attenuated power laws (up
to an energy which remains to be accurately determined), similar to
the power laws observed in the soft state of BHs.  Such power laws
could be produced by non-thermal Comptonization, i.e.  Comptonization
on non-thermal particles (for recent reviews see Poutanen 1999, and
Coppi 1999).  Alternatively, Titarchuk et al.  (1997) have proposed
that these power laws could be produced by bulk-comptonization in a
convergent accretion flow.  The same mechanism seems to be ruled out
for NSs because it produces spectra much harder than that observed
(\alphapl$\la$1, Titarchuk et al.  1996).  BHs with power laws have
typical \alphapl~in the range 2.5-3.0, or similar to NSs in the hard
X-ray band (e.g.  \onee).  Therefore, the indication of a steep hard
X-ray (E$\ga$30 keV) spectrum with \alphapl$\ga$ 2.5 cannot alone be
used to claim that an unknown system contains a NS. Similarly, a hard
X-ray power law spectrum with \alphapl $\la 2.5$ is not unique to BHs
(e.g.  4U0614+091).

\subsubsection{Luminosity differences?}
The idea that BHs and NSs are actually hardly distinguishable by their
broad band spectral shape has led Barret et al.  (1996) to propose a
luminosity criterion (see also Barret \& Vedrenne 1994, Van der Klis
\& Van Paradijs 1994).  They compared the 1-20 keV luminosity (\lx) to
the 20-200 keV luminosity (\lhx) for all secure BHBs (by secure we
mean BHs with mass function estimates indicating a compact object of
mass larger than 3\msol, hereafter BHBs for BH Binaries), and all NSs
detected up to at least 100 keV. Figure 13 is an updated version of
Fig. 1 in Barret et al.  (1996).  There are now 16 NSs detected at 100
keV with (quasi-)simultaneous coverage in X-rays; all but Cen X-4
detections came within the last 8 years. It includes 6 more NSs
recently detected by either RXTE (PCA+HEXTE) or
\sax~(MECS+LECS+HPGSPC+PDS); namely the millisecond pulsar
SAXJ1808.4-3658 (Gilfanov et al.  1998, Heindl \& Smith 1998), the
dippers 4U1915-05 (Church et al.  1998) and XB1323-619
(Balucinska-Church et al., 1999), SAXJ1748.8-2021 (In't Zand et al.
1999), \slx~and \gs~(this work).  For the two dippers, \lx~and
\lhx~have been computed from their non-dip spectra.  In addition, the
luminosities for \onee~(this work), for 4U0614+091 (\sax~observation,
Piraino et al.  1999), and Aql X-1 (joint nearly simultaneous BATSE \&
ASCA observations, Rubin et al.  1999) have been updated. The figure
also includes GRS1009-45 (Nova Velorum 1993), recently shown to be
another secure BHB (Filippenko et al.  1999).

For both SAXJ1808.4-3658 and 4U1915-05, we have assumed the distance
inferred from X-ray burst studies: 4 and 9.3 kpc respectively (In't
Zand et al. 1998, Yoshida 1992), and for XB1323-619 a distance of 10
kpc (Balucinska-Church et al., 1999). SAXJ1748.8-2021 is located in
the globular cluster NGC6440 whose distance is estimated to be
$8.5\pm0.5$ kpc (Ortolani et al. 1994).  For GRS1009-45 we have
estimated the distance according to the method described in Barret et
al. (1996).  We determine the radius of the secondary assuming that it
fills its Roche lobe (the radius is then given by the orbital period
\porb=6.86 hours, and its mass, which we take to be 0.5 \msol).  For a
K7 secondary, from the absolute visual flux, we determine the absolute
visual magnitude $M_v\sim 8.2$, according to the values tabulated by
Popper (1980).  Finally, using the apparent dereddened magnitude
corrected for interstellar reddening and for the contribution (fdisk)
by the accretion disk to the continuum flux ($ V_{ quies}=21.4-21.9$,
E(B-V)=0.2, fdisk=60\%, Shahbaz \& Kuulkers 1998), we obtain a
distance in the range 5.0 to 6.5 kpc.  Shahbaz and Kuulkers (1998)
using an empirical linear relationship between the orbital period, the
optical outburst amplitude magnitude derived a distance of 4 kpc. We
adopt in Fig. 13 a distance of 5 kpc. The X-ray and hard X-ray fluxes
for GRS1009-45 have been taken from Barret et al. (1996) from
observations reported by Sunyaev et al. (1994).

\begin{figure}[!t]
  \vspace*{-0.5cm} \centerline{\psfig{figure=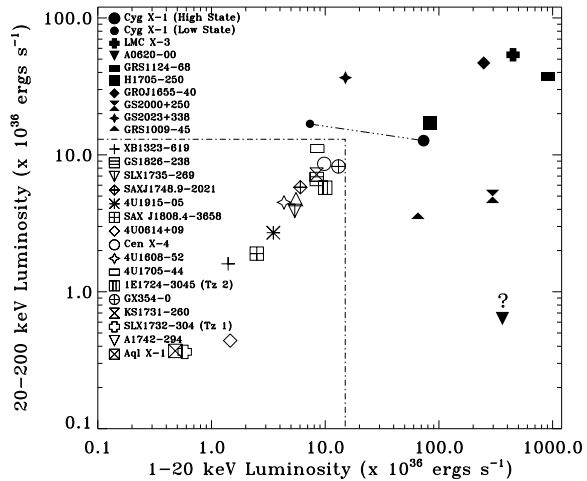,height=8.0cm}}
  \vspace*{-1.0cm}\caption{The hard X-ray (20-200 keV) versus the
    X-ray (1-20 keV) luminosities of Black Hole Binaries (filled
    symbols) and Neutron Star Binaries (open symbols). Only BHs with
    mass functions indicating a mass for the compact object in excess
    of 3\msol~are considered.  For NSs, the luminosities are computed
    from observations in which the source was detected up to at least
    100 keV. For \ks, the luminosities drawn correspond to the SIGMA
    detection (Barret et al. 1992). This is an updated version of Fig.
    1 in Barret et al.  (1996) from which most luminosities can be
    retrieved. The so-called {\it X-ray burster box} is plotted as a
    dot-dashed line. Its boundaries are simply defined as to contain
    all the NSs.}
\label{bhnslum}
\end{figure}

Looking at Fig.  \ref{bhnslum}, one can see that the distinction in
luminosity clearly holds: all NSs lie in the so-called {\it X-ray
  burster box}, whereas all BHs are found outside.  This figure shows
that comparisons of \lhx~and \lx~can be used to distinguish between
BHs and NSs.  First, although the presence of a hard tail is not a
unique signature of BH accretion, the high luminosity in the hard tail
appears to be such a signature: only BHBs are observed with \lhx $\ga$
\lc, where we define \lc~$ \sim 1.5 \times 10^{37}$ \ergs.  Second,
the hard X-ray luminosity of NSs can approach \lc, but only when \lx
$\la$ \lc.  On the other hand BHBs can have both \lx~and \lhx~above
\lc~(although \lhx~may then be substantially lower than \lx).
Finally, the most recent broad band measurements of LS NSs as those
presented in this paper all indicate that these systems have a ratio
\lx/\lhx~in a narrow range of $\sim 1.0-2.0$.  This yields the
correlation between \lx~and \lhx~seen in Fig.  \ref{bhnslum}.  This
suggests that below a luminosity $\sim 2 \times$ \lc, the spectral
shape of a given NS does not depend much on its luminosity.  However,
observation of softer spectra (almost blackbody-like, Asai et al.
1998) from quiescent NS transients imply that this correlation breaks
down somewhere between $\sim 10^{34}$ and $10^{36}$ \ergs.  Another
interesting feature of Fig.  13 is the fact that the BHBs (those
lacking an ultra soft component) lie roughly on an extrapolation of
the sequence formed by the NSs points.  This again points to the
underlying similarity of their emission sources and basic accretion
geometries.

\section{Summary}

We have reported RXTE timing and spectral observations of four type I
X-ray bursters.  The results presented here nicely illustrate that
beside its unprecedented timing capabilities, RXTE has also sufficient
sensitivity for detailed spectral studies.  Our spectral fitting has
revealed that the broad band spectra of low state NSs are often more
complicated than previously thought.  In particular, we have detected
a Compton reflected component plus an iron \kalpha~line in two
systems.  In both cases, the data indicate that the most likely site
for the reflector is a cool accretion disk, which is truncated
somewhere to offer a relatively small solid angle to the irradiating
source, which could be a quasi-spherical hot advection dominated inner
accretion flow.  Similar behavior is observed among BHs leading to
very similar conclusions.  In addition, within the limitations of the
PCA calibration uncertainties at low energies, we have detected a soft
component, which is equally well described by a multi-color disk
blackbody or a blackbody.  In our picture, this soft component most
probably originates from the truncated accretion disk.

At higher energies, the HEXTE sensitivity is sufficient to accurately
locate the energy cutoff in the low state hard X-ray spectra of these
systems.  When combined with PCA data, we have shown that the spectral
steepness of the hard tails observed by SIGMA/BATSE was not intrinsic,
but was rather due to the presence of a high energy cutoff ($ \sim
50-80$ keV).  The cutoff power law spectral shape observed strongly
suggests that thermal Comptonization is the dominant emission
mechanism in these systems.  The Comptonization process would take
place in an optically thin hot boundary layer merged with the central
advective corona, located between the truncated accretion disk and the
NS surface.

Of the most recent criteria that have been proposed to distinguish BHs
from NSs (out of quiescence), two appear consistent with all the
available data. The first one states that for NSs displaying thermal
Comptonization, they seem to be unable to achieve electron temperature
as high as BHs, and hence Comptonization with \kte~$\ga$~50 keV would
be a BH signature.  This criterion is weakened however by the fact
that there are some NSs that do no display high energy cutoffs in
their hard X-ray spectra.  The second one is a luminosity based
argument and claims that only BHs are capable of emitting bright hard
X-ray tails with luminosity larger than $ \sim 1.5 \times 10^{37}$
\ergs.

The RXTE archive contain a wealth of observations of NSs and BHs in
different luminosity states, and represent a very uniform data set,
that should be used to test the picture proposed in this paper. It is
also clear that the same data can be used to challenge the
observational criteria discussed above to distinguish BHs from NSs.

\section{Acknowledgments}
This research has made use of data obtained through the High Energy
Astrophysics Science Archive Research Center operated by the NASA
Goddard Space Flight Center.  GKS is grateful to CESR for its
hospitality during the period of this work.  DB wishes to thank the
organizers of 1999 ITP Black Hole Program (R. Blandford, D. Eardley,
and J.P. Lasota) and the participants (R. Taam, S. Kato, A. Zdziarski,
P. Coppi and O. Blaes) and for their hospitality at the Institute of
Theoretical Physics where part of this work was completed.  DB
acknowledges many exciting discussions with all of them.  This
research was thus supported in part by the National Science Foundation
under Grant No.  PHY94-07194.  DB also thanks the organizers of the
Aspen Center for Physics summer workshop on ``X-Ray Probes of
Relativistic Effects near Neutron Stars and Black Holes'' (J.E.
Grindlay, P. Kaaret, W. Kluzniak, F. Lamb, M. Nowak, W. Zhang) for
inviting him, and for their hospitality during the workshop.

The authors are grateful to J. Swank and K. Jahoda for discussions
about the PCA calibrations, to L. Titarchuk for useful discussions
during his stay at CESR, to A. Zdziarski for providing us with the
GINGA spectrum of \gs, to J. Poutanen for the supply of his
relativistic Comptonization model (\comps), and finally to P. Kaaret
for sending us in advance a copy of the paper Piraino et al.  (1999)
on 4U0614+09.

This paper greatly benefited from many useful comments by A. Bazzano,
P.F. Bloser, E.C. Ford, M. Guainazzi, P. Kaaret, W. Kluzniak, J.P.
Lasota, J.E. McClintock, J. Poutanen, R. Taam, M. Van der Klis, A.
Zdziarski, S.N. Zhang, and W. Zhang.

Finally, we are thankful to Dimitrios Psaltis, the referee, for many
insightful and thoughtful comments on this paper.

 \end{document}